\documentclass[fleqn,usenatbib]{mnras}

\usepackage{newtxtext,newtxmath}

\usepackage[T1]{fontenc}
\usepackage{ae,aecompl}

\usepackage{graphicx}	
\usepackage{amsmath}	
\usepackage{amssymb}	
\RequirePackage{esvect} 
\usepackage{multirow}   
\usepackage{comment}    
\usepackage{threeparttable} 
\usepackage[dvipsnames]{xcolor} 
\usepackage{verbatim} 
\usepackage{courier} 

\usepackage{color}
\definecolor{red}{rgb}{1.0,0.2,0.2}
\definecolor{green}{rgb}{0.2,1.0,0.2}
\usepackage{soul} 

\title[Probabilistic Fringe-fitting]{A probabilistic approach to phase calibration: I. Effects of source structure on fringe-fitting}

\author[Natarajan et al.]{Iniyan Natarajan$^{1}$\thanks{E-mail: iniyannatarajan@gmail.com},
Roger Deane$^{2,1}$,
Ilse van Bemmel$^{3}$,
Huib Jan van Langevelde$^{3,4}$,
\newauthor
Des Small$^{3}$,
Mark Kettenis$^{3}$,
Zsolt Paragi$^{3}$,
Oleg Smirnov$^{1,5}$,
Arpad Szomoru$^{3}$\\
$^{1}$Centre for Radio Astronomy Techniques and Technologies, Department of Physics and Electronics, Rhodes University, \\Grahamstown 6140, South Africa\\
$^{2}$Department of Physics, University of Pretoria, Hatfield, Pretoria, 0028, South Africa\\
$^{3}$ Joint Institute for VLBI ERIC, Oude Hoogeveensedijk 4, 7991 PD Dwingeloo, Netherlands\\
$^{4}$ Sterrewacht Leiden, Leiden University, Postbus 9513, 2300 RA Leiden, Netherlands\\
$^{5}$ South African Radio Astronomy Observatory, Observatory 7925, Cape Town, South Africa
}

\date{Accepted XXX. Received YYY; in original form ZZZ}

\pubyear{2020}

\begin{document}
\label{firstpage}
\pagerange{\pageref{firstpage}--\pageref{lastpage}}
\maketitle

\begin{abstract}
We propose a probabilistic framework for performing simultaneous estimation of source structure and fringe-fitting parameters in Very Long Baseline Interferometry (VLBI) observations. As a first step, we demonstrate this technique through the analysis of synthetic short-duration Event Horizon Telescope (EHT) observations of various geometric source models at 230 GHz, in the presence of baseline-dependent thermal noise. We perform Bayesian parameter estimation and model selection between the different source models to obtain reliable uncertainty estimates and correlations between various source and fringe-fitting related model parameters. We also compare the Bayesian posteriors with those obtained using widely-used VLBI data reduction packages such as \textsc{casa} and \textsc{aips}, by fringe-fitting 200 Monte Carlo simulations of each source model with different noise realisations, to obtain distributions of the Maximum A Posteriori (MAP) estimates.
We find that, in the presence of resolved asymmetric source structure and a given array geometry, the traditional practice of fringe-fitting with a point source model yields appreciable offsets in the estimated phase residuals, potentially biasing or limiting the dynamic range of the starting model used for self-calibration.
Simultaneously estimating the source structure earlier in the calibration process with formal uncertainties improves the precision and accuracy of fringe-fitting and establishes the potential of the available data especially when there is little prior information. We also note the potential applications of this method to astrometry and geodesy for specific science cases and the planned improvements to the computational performance and analyses of more complex source distributions.
\end{abstract}

\begin{keywords}
techniques: interferometric -- techniques: high angular resolution -- methods: data analysis -- methods: statistical
\end{keywords}



\section{Introduction}
\label{sec:intro}
Interferometry is concerned with estimating the spatial coherence function of electromagnetic fields measured between two different locations. A correlator must cross-correlate the signals from different stations while ensuring that they correspond to the same incoming wavefront instance. VLBI observations are similar to those made using connected-element interferometers insofar as the aim is to obtain this coherence function, but the measured visibility phases are less well-behaved in VLBI due to a number of reasons. The atmospheric conditions at individual stations are different, leading to different atmospheric propagation delays that are uncorrelated \citep[][Chapter~9]{TMS2017}. The use of independent frequency standards introduces another source of systematic uncertainty in determining the correct timestamps at which to cross-correlate the signals from different stations. Moreover, the rotation of the Earth causes the stations to move at different speeds with respect to the source, resulting in different Doppler shifts of the incoming wave, which introduces a time-variable delay \citep[][Chapter~22]{Syn1999}. Finally, the presence of complex source structure also contributes to the variations in the phases of the measured visibilities.

A range of geometric, atmospheric, and instrumental effects are accounted for by the \textit{correlator model}, a geometrical model that is applied to correct the visibility phases. This model is not perfect and contains residual errors in phases, that may vary appreciably with frequency and time. Various systematics such as errors in the source or antenna positions, offsets in the clock epoch, and errors in modelling the turbulence in the troposphere, especially at high frequency \citep[][Chapter~28]{Syn1999}, contribute to these residual phases, and considerably reduce the time and frequency interval over which the measured visibilities can be averaged without a net loss of amplitude (\textit{coherence time}), limiting the signal-to-noise ratio (SNR) of the data \citep{schwabcotton1983}.
Correcting for these phase residuals and their slopes with frequency and time (known as \emph{delay residuals} and \emph{rate residuals} respectively)
is known as \textit{fringe-fitting} or \textit{fringe-searching}, which
allows one to average data over larger intervals of time and frequency, thereby increasing the coherence time and bandwidth, and improving the sensitivity.

Fringe-fitting is performed as one of the first steps after correlation and the resulting corrected visibilities are used as inputs for subsequent calibration procedures, such as self-calibration. Hence, studying any correlations between spatially-resolved source structure and fringe-fitting parameters and propagating reliable uncertainties further down to the data analysis steps where the astrophysical parameters are estimated is useful, especially since fringe-fitting is often performed with the assumption that the source is unresolved. Using the wrong source model during fringe-fitting can introduce false symmetry in the source structure, which may affect the results significantly when analysing strongly averaged data \citep[e.g.][]{wielgus2019, eht3}. This may also be relevant to other science cases such as the asymmetric source structure observed in Global mm VLBI Array (GMVA) observations of Sgr A* \citep{issaoun2019}, compact and extended structures observed in VLBI observations of gravitationally lensed radio sources \citep{spingola2019}, and highly accurate source position estimates required for performing VLBI astrometry for the Bulge Asymmetries and Dynamic Evolution (BAaDE) targets \citep{huib2019}.

Incorporating a priori knowledge of the relevant parameters for self-calibration was first advocated by \citet{cornwellwilk1981} and a similar approach for fringe-fitting was advocated by \citet{schwabcotton1983}. Reliable uncertainty propagation for calibration parameters has been discussed within the framework of Information Field Theory (IFT) by \citet{ensslin2009,ensslin2014, ensslin2018}, and implemented in algorithms such as \textsc{resolve} \citep{resolve2016}. \citet{biro2015} discuss simultaneous estimation of source and instrumental parameters and model selection between partially-resolved source structures on Westerbork Synthesis Radio Telescope (WSRT) simulations within a Bayesian framework.
In \citet{iniyan2017}, we undertook a full Bayesian analysis of European VLBI Network (EVN) observations of the blazar J0809+5341 to prove the existence of extended jet structure in the presence of residual station gain amplitudes after self-calibration. 

In this first of an intended series of papers on the application of probabilistic techniques to the wider problem of phase calibration in VLBI, we study the mutual effects of source structure and some fringe-fitting related parameters on each other, using short duration synthetic EHT observations with sparse uv-coverage\footnote{Throughout this paper, we use the terms \emph{simulated} and \emph{synthetic} interchangeably, as both are used in the relevant VLBI literature \citep[e.g.][]{eht4}.}. Of the parameters relevant to fringe-fitting, we simulate the phase and delay residuals; in actual observations, the phases also vary with time (\emph{rate residuals}) and must be accounted for while fringe-fitting. However, in this proof-of-concept presentation of simultaneous source structure estimation and fringe-fitting, we perform these experiments only with simulated phase and delay residuals, since rate residuals are not significant for short duration snapshot observations such as these. We also assume that accurate a priori amplitude calibration has been performed on the data and do not consider gain amplitudes. Finally, we consider the presence of only one spectral window (subband) in this paper.

This work is organised as follows. In Section~\ref{sec:fringefitting}, we briefly discuss the theory behind fringe-fitting approaches used in current software packages. In Section~\ref{sec:bayesfringe}, we present a probabilistic fringe-fitting formalism that simultaneously estimates the source-related parameters. In Sections~\ref{sec:ffeht} and \ref{sec:disc}, we apply this method to synthetic snapshot EHT observations of partially-and-fully-resolved symmetric and asymmetric geometric source models, and compare our results with those of other calibration packages such as \textsc{casa}\footnote{\url{www.casa.nrao.edu}.} and \textsc{aips}\footnote{\url{www.aips.nrao.edu}.}. Finally, we outline ongoing efforts to update this framework to handle an expanded parameter space involving more complex source models and time-varying phase, delay, and rate residuals simultaneously, and the projects that we subsequently propose to undertake in Section~\ref{sec:future}.

\section{Fringe-fitting}
\label{sec:fringefitting}
\subsection{Terminology}
\label{subsec:terminology}
There appears to be confusion in the relevant literature about the nomenclature used to describe the quantities that are estimated during fringe-fitting. Throughout this paper, we consistently refer to the first order phase residual terms ($\psi_{0_p}$) as \emph{phase residuals}; the derivative of the phase with respect to frequency ($\tau_p$) as \emph{delay residual} and the derivative of the phase with respect to time ($r_p$) as \emph{rate residual}.

\subsection{Background}
\label{sec:ffbg}
The significant phase variations present in VLBI data lead to the loss of amplitude of the measured visibilities when averaged (\textit{decoherence}), which fringe-fitting aims to minimise. These variations, to first order, manifest as residual phases, and phase slopes with frequency and time for ground-based VLBI arrays. Let $\widetilde{\mathrm{V}}_{pq}(t_m, \nu_n)$ be the measured visibility corresponding to baseline $pq$ at time $t_m$ and frequency $\nu_n$. This is related to the true visibility $\mathrm{V}_{pq}(t_m, \nu_n)$ as
\begin{equation}
\label{eq:ff1}
\widetilde{\mathrm{V}}_{pq}(t_m, \nu_n) = g_p(t_m, \nu_n) \bar{g}_q(t_m, \nu_n) \mathrm{V}_{pq}(t_m, \nu_n) + \varepsilon_{pqmn}\quad,
\end{equation}
where $\varepsilon_{pqmn}$ is the additive noise term arising due to thermal noise;
$g_p$ are complex-valued functions that incorporate antenna-based effects and can be expressed as
\begin{equation}
\label{eq:ff2}
g_p(t,\nu) = |g_p(t,\nu)|\mathrm{e}^{i\mathrm{\psi}_p(t,\mathrm\nu)}\quad.
\end{equation}
Assuming that the amplitudes $|g_p|$ and $|\mathrm{V}_{pq}|$ vary slowly enough that they
are constant over the time and frequency averaging intervals, we can write to first-order \citep{schwabcotton1983}
\begin{equation}
\begin{split}
\label{eq:ff3}
\widetilde{\mathrm{V}}_{pq}(t_m,\nu_n) \simeq |g_p||g_q|\mathrm{V}_{pq}(t_0,\nu_0)\exp(i[(\mathrm{\psi}_p-\mathrm{\psi}_q)(t_0,\nu_0)]) \\
\times\exp\left(i\left[\frac{\partial(\mathrm{\psi}_p-\mathrm{\psi}_q+\phi_{pq})}{\partial t}\Bigr|_{\substack{(t_0,\mathrm{\nu}_0)}}(t_m-t_0)\right.\right.\\
\left.\left.+\frac{\partial(\mathrm{\psi}_p-\mathrm{\psi}_q+\phi_{pq})}{\partial \nu}\Bigr|_{\substack{(t_0,\mathrm{\nu}_0)}}(\nu_n-\nu_0)\right]\right)\quad,
\end{split}
\end{equation}
where $\phi_{pq} \equiv \arg \mathrm{V}_{pq}$. Here $t_0$ and $\nu_0$ are the time and frequency
relative to which the calculations are carried out. It must be noted that using a fixed reference frequency across the band is valid only for moderate bandwidths as a first-order approximation, which will need to be revised for wide-band fringe-fitting. Here, we adopt this formalism from \citet{schwabcotton1983} which forms the basis of fringe-fitting in both \textsc{casa} and \textsc{aips}. The derivative of phase with respect to time
\begin{equation}
\label{eq:ff4}
r_{pq} \equiv \frac{\partial(\mathrm{\psi}_p-\mathrm{\psi}_q+\phi_{pq})}{\partial t}\Bigr|_{\substack{(t_0,\mathrm{\nu}_0)}}
\end{equation}
is the expression for the rate residual, and the derivative of phase with respect to frequency
\begin{equation}
\label{eq:ff5}
\tau_{pq} \equiv \frac{\partial(\mathrm{\psi}_p-\mathrm{\psi}_q+\phi_{pq})}{\partial \nu}\Bigr|_{\substack{(t_0,\mathrm{\nu}_0)}}
\end{equation}
quantifies the delay residual, estimated relative to $t_0$ and $\nu_0$. These values are estimated by Fourier-transforming the per-baseline visibilities from the time and frequency domain to the delay and rate residual domain and finding its maximum (which would occur at $r_{pq}$ and $\tau_{pq}$). The position of this peak is used to define the phase centre of the observation, a fixed point in the sky relative to which the delay and rate residuals are estimated. The value of the function at this maximum (dependent on $t$ and $\nu$), gives the phase-corrected estimates of the  visibility function in the time-frequency domain \citep{cotton1995}. It must be remembered that these estimates are only approximations to the fringe solutions, since the fringe-rate correction is constant over all frequencies. These can then be averaged coherently over time and frequency, to the extent that the first-order model in equation (\ref{eq:ff3}) is valid. It is important to note that the derivatives of the antenna phases $\mathrm{\psi}$ and those of the true visibility phases $\phi$ cannot be separated by this method.

Global fringe-fitting is performed to separate the antenna-based components in equation (\ref{eq:ff3}) and estimate them simultaneously for all stations using the data corresponding to all baselines. This is especially useful in cases where the SNR is low, which is often the case with VLBI observations. If the source is resolved, then a model of the source $\mathrm{V}^M_{pq}$ approximating the true visibilities $\mathrm{V}_{pq}$ is needed to enable this separation. If the source is (a) sufficiently compact so the 2-D FFT to the delay and rate residual domain has a well-defined peak and (b) the visibility function does not change significantly during the integration period \citep{alefporcas1986}, then the default assumption of the source being point-like is made. For more elaborate source models, in theory, an initial image of the source can be used to enable the separation of baseline-based and antenna-based phase components \citep{schwabcotton1983}.
Global fringe-fitting is related to phase self-calibration \citep{selfcal1978,readhead1980} in that both procedures aim to minimise the difference between the observed and model visibility phases. In addition to phase residuals, fringe-fitting also estimates their time and frequency slopes, and is always performed before averaging in order to preserve coherence. 

Contemporary fringe-fitting algorithms such as the one employed by the \textsc{casa} task \texttt{fringefit} \footnote{\url{https://casa.nrao.edu/casadocs-devel/stable/global-task-list/task_fringefit/about}.}, perform baseline-based fringe-fitting before globalising 
the solutions to all baselines through a least-squares method \citep[e.g.][]{janssen2019}. This can be done, for instance, by performing baseline-based fringe-fitting with a subset of the more sensitive stations in the array, and letting the result serve as a starting point to a later global fringe-fitting step. A baseline-based approach that finds the maximum in a three-dimensional space of multi-band delay, single-band delay, and rate, is employed by the \texttt{fourfit} task \citep{fourfit} in the Haystack Observatory Postprocessing System (\textsc{hops})\footnote{\url{www.haystack.mit.edu/tech/vlbi/hops.html}.} \citep{whitney2004}.
A pipeline based on \textsc{hops} that performs additional global fringe-fitting for delay and rate residuals \citep{blackburn2019} is one of the primary fringe-fitting tools used to process data from the EHT, along with \textsc{casa} and \textsc{aips} \citep{eht3}, the tools we compare our results with in this work.

\section{A RIME-based Probabilistic Approach}
\label{sec:bayesfringe}
We adopt a fully Bayesian approach to simultaneously estimate the source structure and the phase and delay residuals. This approach has the significant advantage that the prior assumptions (or biases) are quantified and propagated into the final visibility or image-plane analysis \citep[e.g.][]{iniyan2017}. A natural outcome of this process is that we obtain reliable uncertainty estimates of and degeneracies between the various model parameters. A \emph{model} or a \emph{hypothesis} in our context, includes both source and instrumental parameters.

We use the Radio Interferometry Measurement Equation (RIME) formalism for modelling visibilities, which was originally developed for radio polarimetry by \citet{hbs1996} and extended to incorporate direction-dependent effects by \citet{oms2011}.
The RIME expresses the relationship between the source and instrumental characteristics in the form of $2\times 2$ complex matrices known as \emph{Jones} matrices \citep{jones1941}, which enables us to model the desired source properties and propagation path effects in parametric form.
The generic RIME for a sky composed of multiple discrete sources is given by
\begin{equation}
\label{eq:rimemult}
\mathrm{V}_{pq} = \sum_{s} J_{sp}\, \mathrm{B}_s\, \, J_{sq}\! ^H\quad,
\end{equation}
where ${\rm V}_{pq}$ denotes the visibility corresponding to baseline $pq$, $J_{sp}$ represents the $2\times 2$ cumulative Jones matrix corresponding to antenna $p$ in the direction of source $s$, and $B_s$ is the brightness matrix corresponding to the full polarisation output of the correlator.
Each propagation path effect may be assigned its own Jones matrix, with the unknown or inseparable effects being subsumed into a generic Jones term. 

The RIME for fringe-fitting is constructed as follows. In our analyses for this paper, we simulate only unpolarised sources with flat spectra. Hence, the brightness matrix for a source with flux density $S_{\nu}$ at frequency $\nu$ is given by
\begin{equation}
\label{eq:bmatfd}
\mathbfss{B} = \begin{pmatrix} S_{\nu} && 0 \\ 0 && S_{\nu} \end{pmatrix}\quad.
\end{equation}
This signal undergoes a linear transformation represented by the phase delay matrix $K$, representing the phase difference ($\kappa_p$) between the waves received by antenna $p$
located at $\mathbf{u}_p = (u_p, v_p, w_p)$ relative to $\mathbf{u}=0$:
\begin{equation}
\label{eq:kjones}
K_p = \mathrm{e}^{-i\kappa_p} \equiv \mathrm{e}^{-i\kappa_p} \begin{pmatrix} 1 && 0 \\ 0 && 1 \end{pmatrix}\quad.
\end{equation}
Since the K-Jones term exists even under ideal conditions with no propagation path effects, we define the \textit{source coherency} matrix for convenience as \citep{oms2011}
\begin{equation}
\label{eq:soucoh}
\mathbfss{X}_{pq} = K_p\, \mathbfss{B}\, K_q^H = \mathbfss{B}\mathrm{e}^{-i\kappa_{pq}}\quad.
\end{equation}
The E-Jones terms for primary beams are taken to be unity
($E_{sp}\equiv 1$) owing to the small FoVs often used for VLBI data analysis\footnote{For a discussion of the effects of E-Jones terms on EHT observations in the presence of antenna pointing errors, see \citet{tariq2017}.}.
We construct a scalar phase-matrix $G$ that affects both polarisation axes equally:
\begin{equation}
\label{eq:bayesfringe2}
G_p = |g_p| \exp(i[\Delta\mathrm\psi_p(t,\nu)]) \begin{pmatrix} 1 && 0 \\ 0 && 1 \end{pmatrix}\quad,
\end{equation}
where $\Delta\mathrm\psi_p$ incorporates two independent terms that affect the visibility phases: the station-based phase residuals $\psi_{0_p}$ and delay residuals $\tau_p$, relative to a reference antenna and frequency $\nu_{\rm ref}$:
\begin{equation}
\begin{aligned}
\label{eq:bayesfringe3}
\Delta\mathrm\psi_p(t_m,\nu_n) &= \mathrm\psi_{0_p} + \frac{\partial\mathrm\psi_p}{\partial\nu}(\nu_n-\nu_{\rm ref})\quad, \\
&= \mathrm\psi_{0_p} + \tau_p(\nu_n-\nu_{\rm ref})\quad.
\end{aligned}
\end{equation}
To keep the execution times (on a small compute server) manageable, for the purposes of this paper we restrict the number of model parameters relevant to fringe fitting, and simulate only the phase and delay residuals. More information on how we propose to expand this framework can be found in Section~\ref{sec:future}.
Including the above terms, equation (\ref{eq:rimemult}) becomes:
\begin{equation}
\label{eq:bayesfringe1}
\mathrm{V}_{pq}(t_m,\nu_n) = G_p(t_m,\nu_n) \left( \sum_{s} \mathrm{X}_{spq}(t_m,\nu_n) \right) G_q^H(t_m,\nu_n)\quad.
\end{equation}

For fringe-fitting, we are interested only in the phases of the $G_p$ terms and set the gain amplitudes to unity ($|g_p|\equiv1$). In practice, these terms are always present and uncertainties in their estimates also affect the estimated source structure \citep[e.g.][]{iniyan2017}.

Having built this formalism for forward-modelling, we perform Bayesian inference on the synthetic data. Assuming a model (or hypothesis) $H$ to be true, we estimate its parameters, $\Theta$, by fitting them to the visibilities (data, $D$)\footnote{A more detailed introduction to Bayesian inference may be found in \citet[][and references therein]{iniyan2017}. Here we give a short review relevant to our analysis.}:
\begin{equation}
\label{eq:bayesparest}
\mathcal{P}(\Theta|D,H) = \frac{\mathcal{P}(\Theta|H)\, \mathcal{P}(D|\Theta,H)}{\mathcal{P}(D|H)}\quad,
\end{equation}
where $\mathcal{P}(\Theta|H)$ is called the \textit{prior} probability distribution, which encodes our beliefs about the parameters prior to the analysis of the data. $\mathcal{P}(\Theta|D,H)$ is the \textit{posterior} probability distribution which describes how the data $D$ modify our initial beliefs. $\mathcal{P}(D|\Theta,H) \equiv \mathcal{L}(\Theta|D,H)$ is the \textit{likelihood}, which reflects how the uncertainties in the measurement are distributed. The denominator $\mathcal{P}(D|H)$ is a normalising constant called the \emph{Bayesian evidence} or the \emph{marginal likelihood} that becomes a valuable tool for ranking models while performing model comparison.
Assuming Gaussian noise, which is a good approximation in the high SNR regime \citep[e.g.][]{eht6}, given the observed ($V_{D}$) and
the modelled ($V_M$) visibilities (refer equation \ref{eq:bayesfringe1}), and the uncertainties $\sigma_{k}$ that vary with baseline, the likelihood function for parameter estimation for model $H$ may be written as
\begin{equation}
\begin{aligned}
\label{eq:ourlikelihood}
\mathcal{L}(\Theta|V_D,H) &= \frac{1}{\displaystyle\prod_{k=1}^{2N_{\mathrm{vis}}}\, \sqrt{2\pi \sigma_k^2}} \exp \left( -\frac{\chi^2}{2} \right)\quad, \\
\mathrm{where}\ \chi^2 &= \sum_{k=1}^{2N_{\mathrm{vis}}}\left(\frac{V_{M_k}-V_{D_k}}{\sigma_k}\right)^2\quad,
\end{aligned}
\end{equation}
and $N_{\mathrm{vis}}$ is the total number of complex visibilities. 
The natural logarithm of $\mathcal{L}$, given by
\begin{equation}
\begin{aligned}
\label{eq:logl}
\mathrm{ln}(\mathcal{L}) &= \sum_{k=1}^{2N_{\mathrm{vis}}} \mathrm{ln} \left[ (2\pi\sigma_k^2)^{-1/2} \right] - \frac{\chi^2}{2} \\
&= -\frac{1}{2} \sum_{k=1}^{2N_{\mathrm{vis}}} \mathrm{ln} \left[ 2\pi\sigma_k^2 \right] - \frac{\chi^2}{2}\quad, \\ 
\end{aligned}
\end{equation}
is often used in practice since it is more convenient to work with.
The baseline-dependent per-visibility noise term for one polarisation is computed from the System Equivalent Flux Densities (SEFDs) of the individual stations using the radiometer equation \citep{TMS2017}:
\begin{equation}
\label{eq:varyingsigma}
\begin{aligned}
\sigma_{pq} &= \frac{\mathrm{SEFD}_{pq}}{\sqrt{2\delta\nu\, t_{pq}}}\quad, \\
\mathrm{where}\ \mathrm{SEFD}_{pq} &= \sqrt{\mathrm{SEFD}_p\, \mathrm{x}\, \mathrm{SEFD}_q}\quad, \\
\end{aligned}
\end{equation}
$\mathrm{SEFD}_p$ is the SEFD of station $p$, $\delta\nu$ is the channel bandwidth, and $t_{pq}$ is the integration time for baseline $pq$.

The second level of inference we perform is model selection between different hypotheses.
We use various geometric source models in our analyses, and always compare the results with those obtained with a point source model.
Given hypothesis $H$, and a prior belief in the validity of $H$ given by $\mathcal{P}(H|I)$, where $I$ is any relevant background information, the model
posterior probability may be computed using the evidence obtained from parameter estimation as
\begin{equation}
\label{eq:bayesmodelsel}
\mathcal{P}(H|D,I) \propto \mathcal{P}(D|H,I)\, \mathcal{P}(H|I)\quad.
\end{equation}
Given two models $H_1$ and $H_2$, we may define a model selection ratio between the posteriors of the two models as
\begin{equation}
\label{eq:modelselrat}
\frac{\mathcal{P}(H_1|D,I)}{\mathcal{P}(H_2|D,I)} = \frac{\mathcal{Z}_1}{\mathcal{Z}_2}\, \frac{\mathcal{P}(H_1|I)}{\mathcal{P}(H_2|I)} = B_{12}\, \frac{\mathcal{P}(H_1|I)}{\mathcal{P}(H_2|I)}\quad,
\end{equation}
where $\mathcal{P}(H_1|I)/\mathcal{P}(H_2|I)$ is the ratio of the priors of the two models which we set to unity,
indicating that there is no prior preference for one model over the other. The ratio of the evidences, $B_{12}$, known as the \textit{Bayes factor} \citep{jeffreys1961} then provides the odds in favour of $H_1$; the larger the value of $B_{12}$, the more is $H_1$ preferred over $H_2$. Hence, Bayes factors between two models provide a more comprehensive metric for model comparison than traditional methods. Following \citet{kassraftery1995}, we use twice the natural logarithm of this factor as a measure of how strongly one model is preferred over another (Table~\ref{tab:evidences}).
\begin{table}
\caption{Criteria for model selection. $B_{12}$ denotes the ratio of the evidences between hypotheses $H_1$ and $H_2$ \citep{kassraftery1995}, which measures the relative success of the two models at predicting the data.}
\begin{center}
\begin{tabular}{ c | c | c }
\hline
$2\, \mathrm{ln} (B_{12})$ & $B_{12}$ & Evidence against $H_2$ \\
\hline
$0$ to $2$ & $1$ to $3$ & Not worth more than a mention \\
$2$ to $6$ & $3$ to $20$ & Positive \\
$6$ to $10$ & $20$ to $150$ & Strong \\
$>10$ & $>150$ & Very strong \\
\hline
\end{tabular}
\end{center}
\label{tab:evidences}
\end{table}
An order of magnitude more support by the data in favour of $H_1$ is required to move up a level on this scale \citep{trotta2008}.

\subsection{Software setup}
\label{subsec:swsetup}
We have developed a software package called \textsc{zagros}\footnote{\url{https://github.com/saiyanprince/zagros}.} to perform simultaneous source parameter estimation and fringe-fitting. \textsc{zagros} uses \textsc{codex-africanus}\footnote{\url{https://github.com/ska-sa/codex-africanus}.}, a GPU-based forward-modelling software package (Perkins et al in prep), that speeds up model computation required for every iteration of the likelihood evaluation. \textsc{zagros} can currently handle small datasets that can fit into the memory of an NVIDIA Tesla K20m GPU.

For sampling from the multi-dimensional posterior distribution we use \textsc{dypolychord}, an implementation of the dynamic nested sampling algorithm \citep{Higson2019}, based on the publicly available \textsc{polychord} tool \citep{polychord1, polychord2}. For generating synthetic EHT observations, we use \textsc{meqsilhouette}, a mm-VLBI synthetic data generation package capable of generating model visibilities from parametric and non-parametric sky models and adding various propagation path effects such as tropospheric phase corruption, variable receiver gains, antenna pointing errors, and polarisation leakage \citep[][Natarajan et al in prep]{tariq2017}.

\section{Fringe-fitting and source structure}
\label{sec:ffeht}
The EHT is a network of mm/sub-mm facilities spread
across continents to create a telescope with high angular resolution ($\simeq30$--$10\, 
\upmu$as), with the longest baselines spanning the Earth's diameter \citep{eht2}.
The primary goal of the EHT is to image the gravitationally-lensed photon `ring' or \textit{shadow} around the event
horizons of the supermassive black holes at the centres of the Milky Way (Sgr A*) and the
supergiant elliptical galaxy M87, which have the largest predicted apparent angular diameters that are resolvable by the EHT at 230 GHz \citep[e.g.][]{broderick2009,falcke2013}. 
In the observing run of April 2017, the EHT imaged the black hole shadow of M87 within an asymmetric ring of size $42\pm3\ \upmu$as \citep{eht4,eht6}. At this frequency, the troposphere gives rise to a turbulent component to the delays, which is the major contributor to the decoherence of the visibilities. As the frequency increases, so does the tropospheric absorption, mainly due to the pressure-broadened transition lines of H$_2$O and O$_2$ \citep{carilli1999}. Unlike other chemical components, water vapour mixes poorly in the atmosphere, introducing rapid fluctuations in the measured visibility phases.

Under such difficult observing conditions, it is instructive to study the effects of partially or fully-resolved source structure on fringe-fitting. This is also relevant for self-calibration downstream, as the effects of the early decisions on the initial models will be impossible to remove further down the data analysis process. This is especially true for arrays with few stations such as EHT (given the small mutual visibility windows between stations, even fewer are observing simultaneously).
The critical point in this paper is that the probabilistic approach has the advantage that it can simultaneously estimate the parameters related to source structure and fringe-fitting, and reveal whether they are degenerate, as well as enabling model selection via the Bayesian evidence. This not only minimises the effects that arise from using incorrect source models, but also enables the propagation of phase uncertainties further down in the calibration process. This may not always yield an appreciable difference in the final calibrated data, but may prove highly useful for high-value science targets such as the ones mentioned in Section \ref{sec:intro}.

With the recent introduction of the fringe-fitting task \texttt{fringefit}, \textsc{casa} is being adopted for VLBI data processing \citep[e.g.][]{janssen2019, ilse2019} and hence serves as a good tool to compare our results with. We also perform some comparisons with the \textsc{aips} task \texttt{FRING}\footnote{\url{http://www.aips.nrao.edu/cgi-bin/ZXHLP2.PL?FRING}.} (for the elliptical Gaussian model), which has traditionally been used for fringe-fitting. Since Bayesian methods are intrinsically concerned with probability distributions, we want to compare them to \emph{distributions} of parameter estimates output by conventional methods; since the latter generate only MAP estimates (albeit with bounds), we run them over a set of 200 Monte Carlo simulations of each sky model in the study, each with a different noise realisation, and thus obtain these distributions.

\subsection{Simulations of geometric source models}
\label{subsec:tests}
We consider the following source models in our simulations:
a point source at the centre (PT), a circular Gaussian source (CIRC), an elliptical Gaussian source (ELLIP), and two point sources (2PT). To each source model, phase and delay residuals and baseline-dependent Gaussian thermal noise are introduced.
The EHT stations used for these simulations are the Submillimeter Array (SMA) and James Clerk Maxwell Telescope (JCMT) in Hawai'i, Submillimeter Telescope (SMT) in Arizona, Large Millimeter Telescope (LMT) in Mexico, the Atacama Large Millimeter Array (ALMA) and the Atacama Pathfinder Experiment (APEX) in Chile, and the South Pole Telescope (SPT) (Table~\ref{tab:ehtstations}).
The phase centre of the observations coincides with the coordinates of Sgr$\, $A*, $\alpha_{\rm J2000} =  17^\mathrm{h}45^\mathrm{m}40^\mathrm{s}.04088,\, \delta_{\rm J2000} = -29^{\circ}0'28''.118$. The mock-observations were conducted for 3 minutes with a 2 s integration time, at a frequency of 230 GHz, with a $2.56$ GHz bandwidth divided into $32$ channels.
\begin{table}
\caption{EHT stations participating in the mock-observations.}
\begin{center}
\begin{tabular}{ccc}
\hline
Station & Diameter (m) & Nominal SEFD (Jy)\\
\hline
SMA (SM) & 25 & 6000\\
SMT (AZ) & 30 & 1300\\
LMT (LM) & 32 & 560\\
ALMA (AA) & 25 & 220\\
JCMT (JC) & 15 & 5000\\
SPT (SP) & 10 & 1600\\
APEX (AP) & 12 & 4500\\
\hline
\end{tabular}
\end{center}
\label{tab:ehtstations}
\end{table}
The start time of the observation
was selected for maximum mutual visibility between the stations.

The parametrisation of each model evaluated is shown in Table~\ref{tab:models} and the hyperparameters used in the analyses are shown in Table~\ref{tab:simpriors}.
\begin{table}
 \caption{Models evaluated in this study. Each model is composed of source-related parameters and the station-based phase and delay residuals. The degrees of freedom (DoF) indicates the number of parameters that are free to vary independently (excludes parameters with delta priors, such as the phase and delay residuals corresponding to reference stations; refer text).}
 \label{tab:models}
 \begin{tabular}{ccc}
  \hline
  Model & Parameters/DoF & Parametrisation\\
  \hline
  \multirow{4}{*}{\textbf{CIRC}} & \multirow{4}{*}{24/14} & Flux Density ($S_{\nu}$) \\
  & & Position ($l, m$) \\
  & & Shape ($e_{\rm maj}, e_{\rm min}, \mathrm{PA}$) \\
  & & Phase residuals ($\psi_{0_p}$) \\
  & & Delay residuals ($\tau_p$) \\ \hline
  \multirow{5}{*}{\textbf{ELLIP}} & \multirow{5}{*}{24/16} & Flux Density ($S_{\nu}$) \\
  & & Position ($l, m$) \\
  & & Shape ($e_{\rm maj}, e_{\rm min}, \mathrm{PA}$) \\
  & & Phase residuals ($\psi_{0_p}$) \\
  & & Delay residuals ($\tau_p$) \\ \hline
  \multirow{5}{*}{\textbf{2PT}} & \multirow{5}{*}{24/16} & Flux Density ($S_{\nu1}$,$S_{\nu2}$) \\
  & & Position (($l_1, m_1$), ($l_2, m_2$)) \\
  & & Phase residuals ($\psi_{0_p}$) \\
  & & Delay residuals ($\tau_p$) \\
  \hline
 \end{tabular}
\end{table}
\begin{table}
 \caption{Prior distributions for all the model parameters. All parameters were set uniform priors with the range indicated by the values in the square brackets. For parameters with delta priors, refer text.}
 \label{tab:simpriors}
 \begin{threeparttable}
 \centering
 \begin{tabular}{ll}
  \hline
  Parameter (Units) & Prior distribution \\
  \hline
  $S_{\nu}$ (Jy) & [0, 2] \\
  $l_2$ ($\upmu$as) & [-50, 50]\\
  $m_2$ ($\upmu$as) & [-10, 110]\\
  $e_{\rm maj}$ ($\upmu$as) & [0, 40] \\
  $e_{\rm min}$ ($\upmu$as) & [0, 40] \\
  $\mathrm{PA}$ ($^{\circ}$) & [0, 180] \\
  $\psi_{0_p}$ ($^{\circ}$), where $p\ne\mathrm{LM}$ & [0, 360] \\
  $\tau_p$ (ps), where $p\ne\mathrm{LM}$ & [-200, 200] \\
  \hline
 \end{tabular}
 \end{threeparttable}
\end{table}
In all the models, the position ($l, m$) of the central source was fixed at the phase centre. The position of the secondary source in 2PT $(l_2, m_2)$ is given a uniform prior overlapping with the position of the primary source. CIRC and ELLIP use the same parameters to describe the shape, with CIRC imposing the additional constraints $e_{\rm min} = e_{\rm maj}$ and $\mathrm{PA}=0$. The position angle $\mathrm{PA}$ is set to vary over a range of 180 degrees, to avoid degeneracies between position angles oriented in opposite directions. LM was used as the reference station and hence its phase and delay residuals were set to zero. The centre frequency of the band was chosen as the reference frequency.

\subsubsection{Point Source and Circular Gaussian (PT \& CIRC)}
\label{sss:ptcirc}
We simulate 21 datasets, each with a central circular Gaussian of total flux density 1 Jy, with increasing HPBW from 0 to 20 $\upmu$as, up to about half the size of the point spread function (PSF) of the array. In all \textsc{zagros} runs, we use 400 live points and 100 points for the initial exploratory run of \textsc{dypolychord}, via the \texttt{nlive\_const} and \texttt{ninit} parameters respectively \citep{Higson2019}.
To perform Bayesian model selection on each dataset, we compare the evidences for a circular Gaussian model (CIRC) with that of a point source model (PT). In each case, the correct model is favoured (PT for the 0 $\upmu$as source and CIRC for the rest) with \textit{very strong} Bayesian evidence (Table~\ref{tab:evidences}). As the source gets more resolved, the odds ratio in favour of CIRC increases from $23:1$ (for 1 $\upmu$as Gaussian) to $10^{12}:1$ (for 20 $\upmu$as Gaussian), indicating that the correct model is very strongly preferred. The error in relative natural-logarithmic evidence (the quantity defined in Table~\ref{tab:evidences}) in each case is $\pm 0.7$.
It must be noted that, in actual observations, the minimum resolvable source size depends on the SNR, array configuration and calibration uncertainties, and hence additional calibration steps will be necessary \citep{martividal2012,iniyan2017}.

\begin{figure*}
\centering
 \includegraphics[scale=1.25]{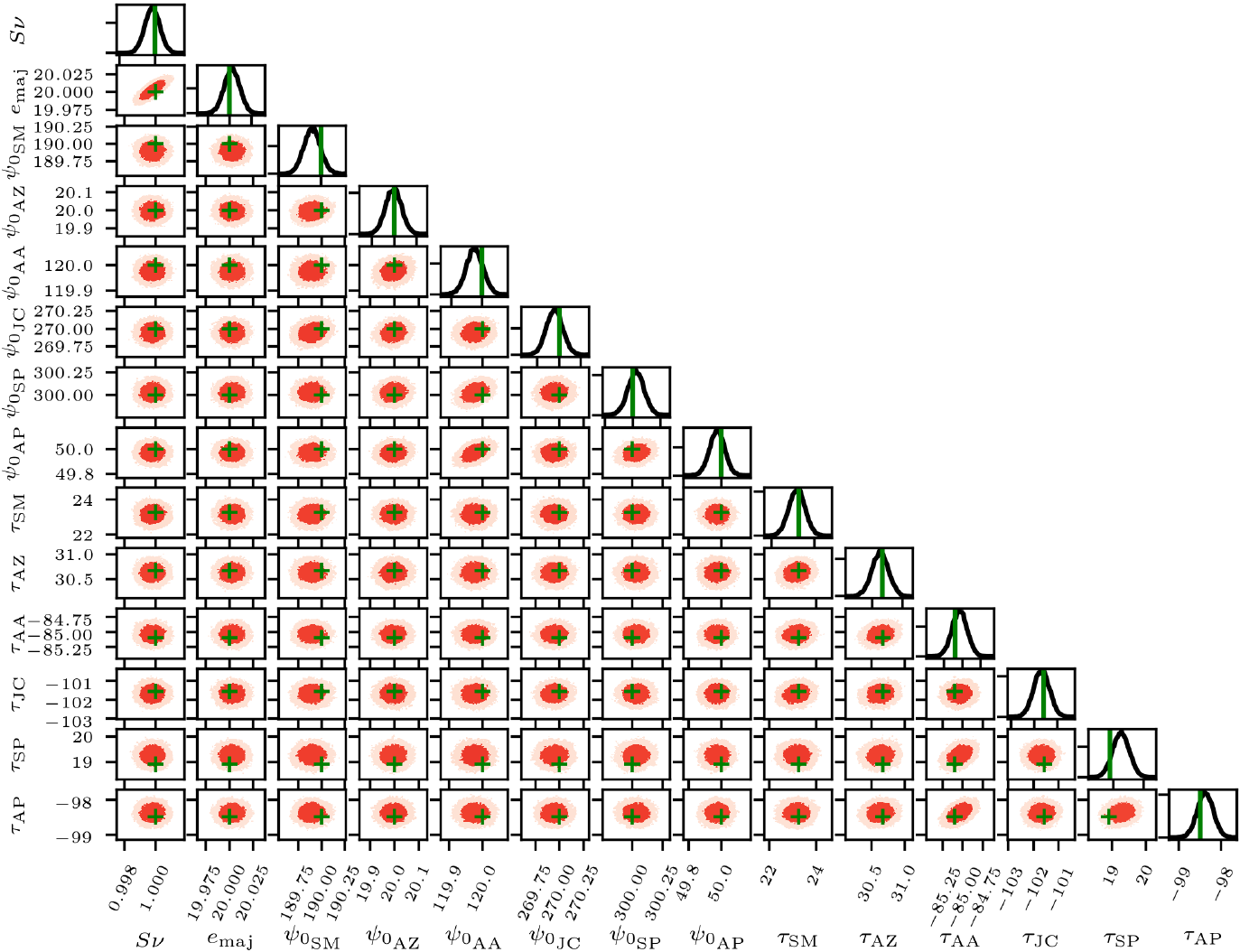}
 \caption[CIRC bayesfringe results]{The 1-D and 2-D posteriors of the parameters of CIRC of size 20 $\upmu$as. The principal diagonal gives the 1-D marginalised posterior distributions of the estimated flux density ($S_{\nu}$), the HPBW of the Gaussian profile ($e_{\mathrm{maj}} = e_{\mathrm{min}})$,
and the station delay residuals $\tau_p$ shifted to zero mean. The 68 (1$\sigma$) and 95 (2$\sigma$) per cent credible regions are indicated by the dark-red and light-red shaded regions respectively. The vertical green lines indicate the ground-truth values used in the simulation. Parameters with delta priors are not shown.}
\label{fig:fwhm20uas_all_triplot}
\end{figure*}

Fig.~\ref{fig:fwhm20uas_all_triplot} shows the results of \textsc{zagros} fringe-fitting on the synthetic data with a 20 $\upmu$as circular Gaussian source at the centre.
We see from the 1-D marginalised posterior plots on the main diagonal, that all the model parameters are estimated to a high precision and accuracy. The 2-D joint posteriors show that there are no significant degeneracies between the source and instrumental parameters in this case, apart from a slight correlation between the delay residual estimates for AA with AP, likely due to their proximity.

\begin{figure}
 \includegraphics[width=\columnwidth, height=0.25\textheight, keepaspectratio]{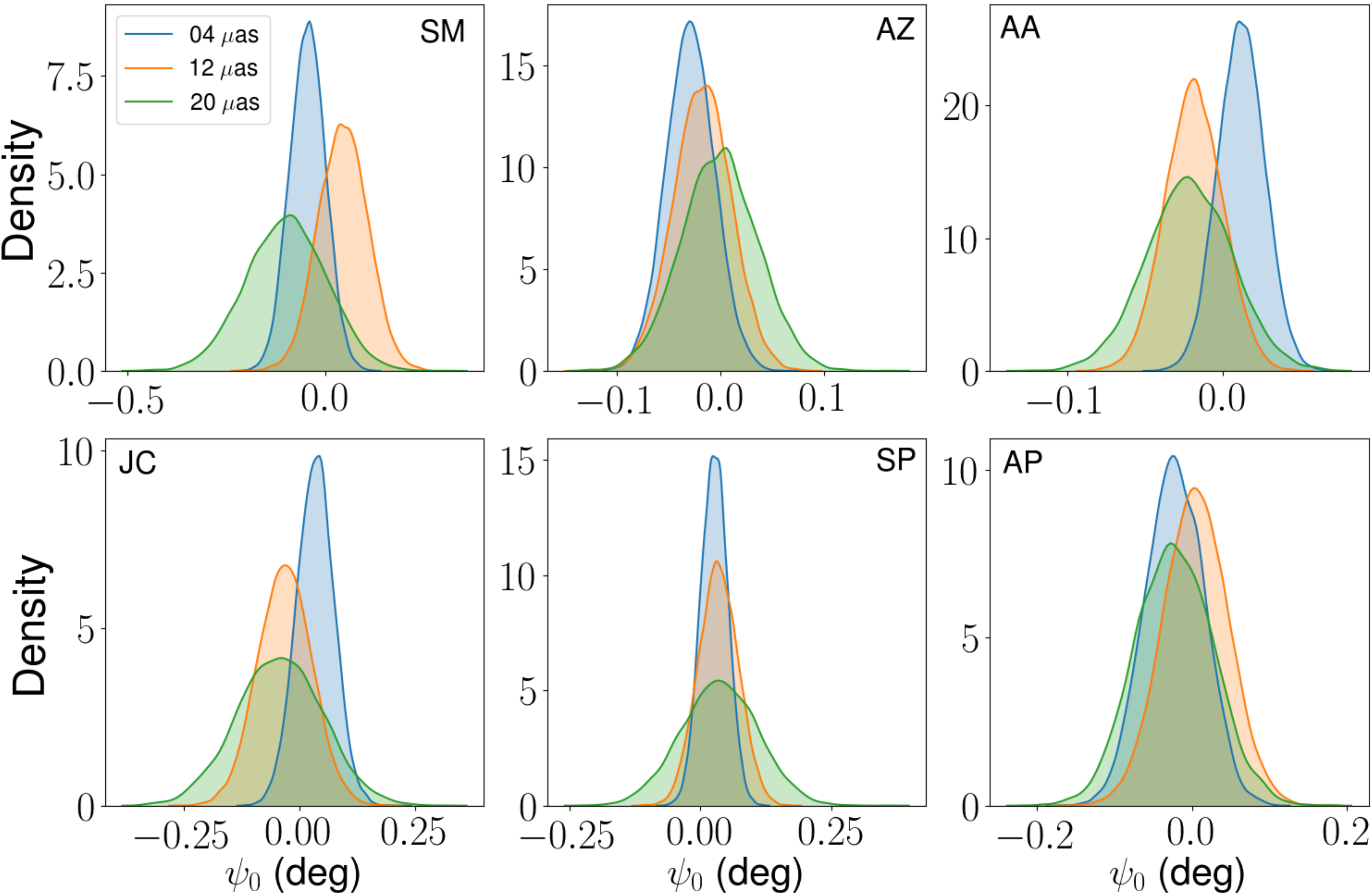}
 \caption[comparesnr-phases]{Posteriors of the phase residuals estimated by \textsc{zagros} for synthetic data generated using CIRC models of three different sizes: 4, 12 and 20 $\upmu$as. The ground-truth values are shifted to zero.}
\label{fig:comparesnr_phases}
\end{figure}

\begin{figure}
 \includegraphics[width=\columnwidth, height=0.25\textheight, keepaspectratio]{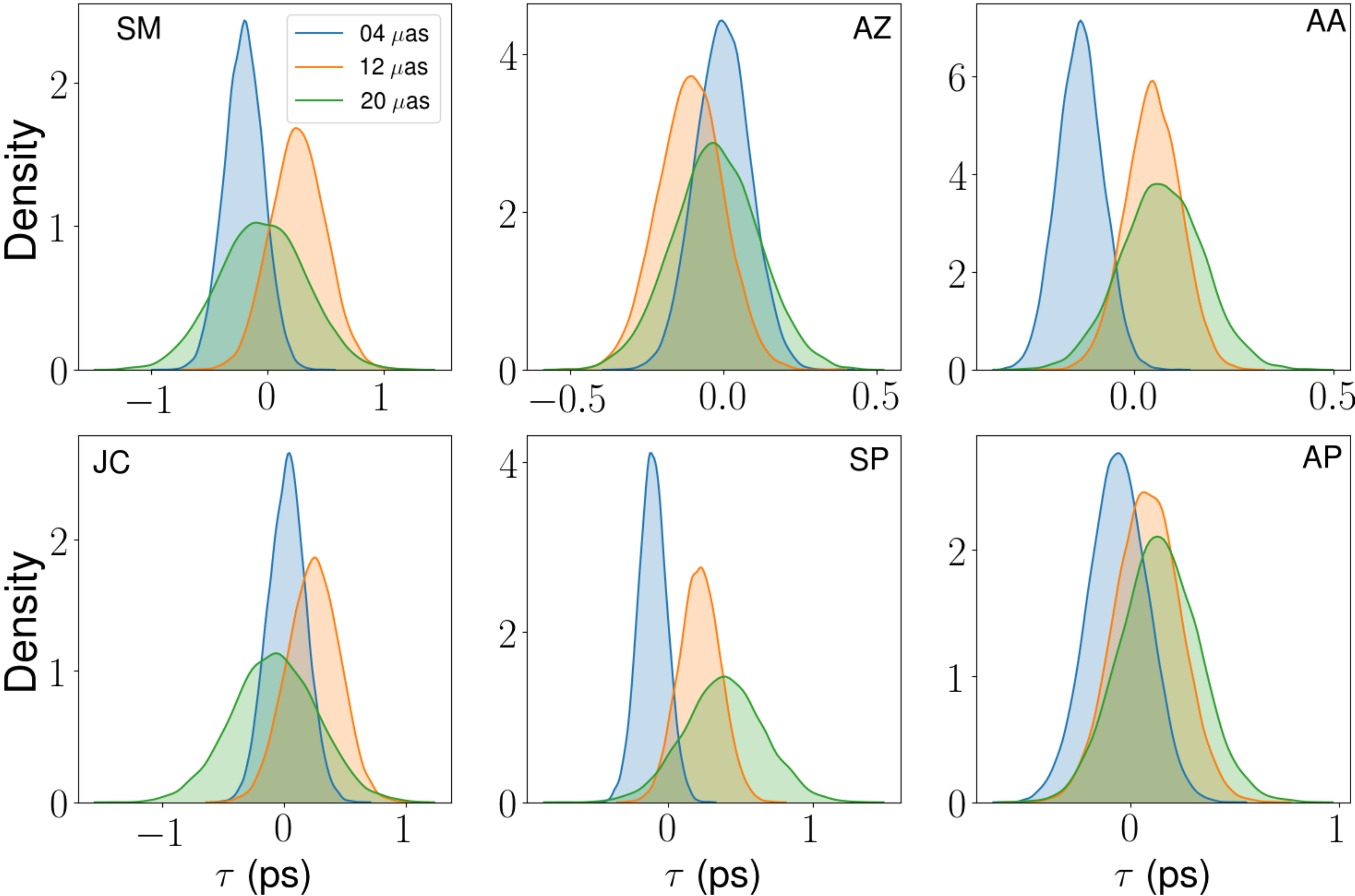}
 \caption[comparesnr-delays]{Same as Fig. \ref{fig:comparesnr_phases}, but for delay residuals.}
\label{fig:comparesnr_delays}
\end{figure}

\begin{figure}
 \includegraphics[width=\columnwidth, height=0.25\textheight, keepaspectratio]{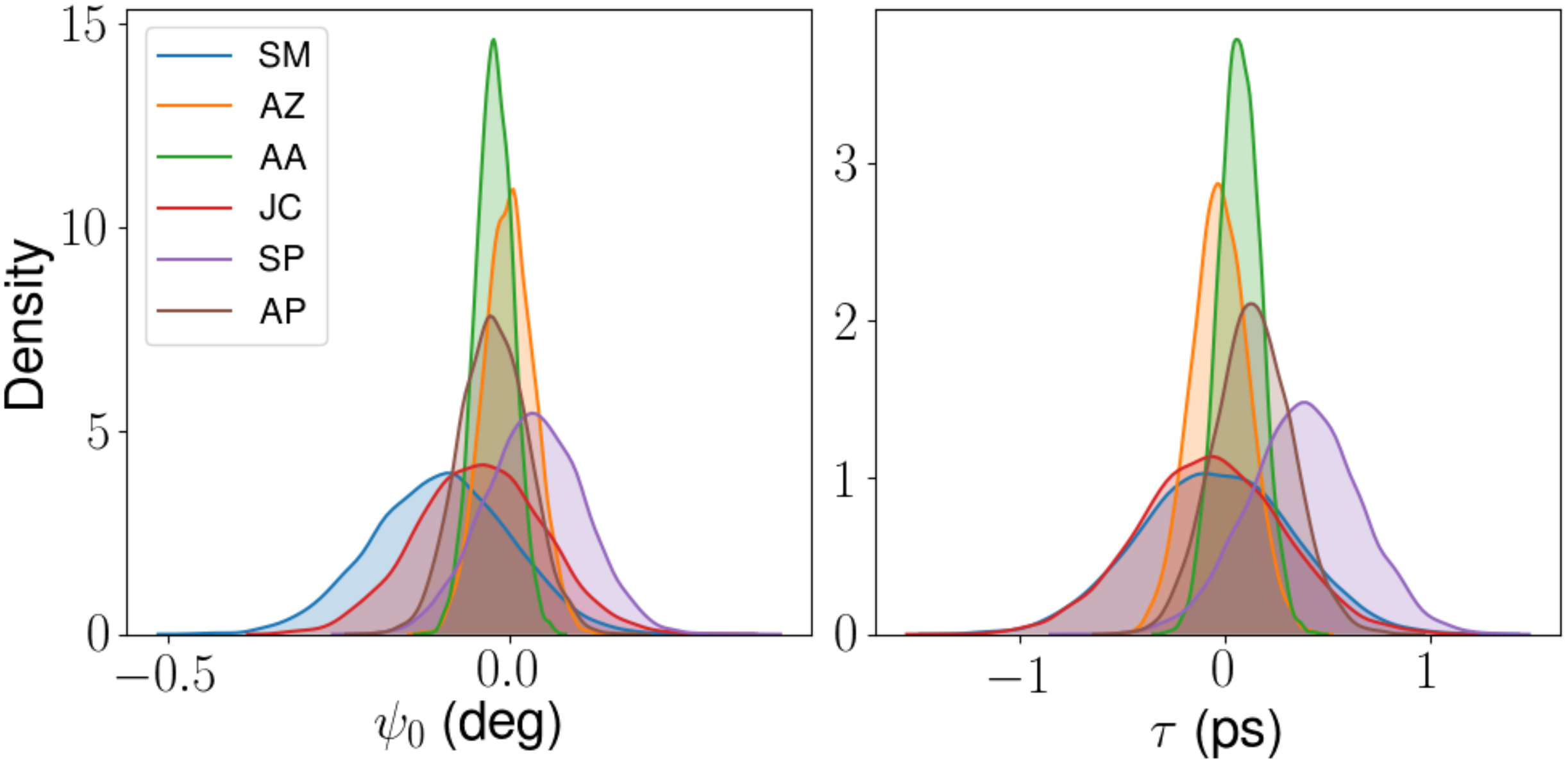}
 \caption[CIRC-phdl-blsnr]{The posteriors of the phase and delay residuals from Fig. \ref{fig:fwhm20uas_all_triplot} (20 $\upmu$as CIRC model) for all antennas. The ground-truth values are shifted to zero.}
\label{fig:CIRC_phdl_blsnr}
\end{figure}

As the radius of the simulated circular Gaussian source increases, the peak flux density goes down which results in decreased SNR of the data. This reduction in SNR contributes to the widening of the posteriors for larger source sizes. Figs. \ref{fig:comparesnr_phases} and \ref{fig:comparesnr_delays} demonstrate this effect for the estimated phase and delay residuals respectively for synthetic data generated with increasingly resolved CIRC models.
In addition, the larger sources are resolved by the longer baselines and the corresponding visibility amplitudes decrease. This effect is demonstrated in Fig. \ref{fig:CIRC_phdl_blsnr} which plots the posteriors of phase and delay residuals for all antennas for the 20 $\upmu$as CIRC model.
The posteriors corresponding to SM, JC, and SP that contribute to the longest baselines are wider than those of the other stations whose baselines are relatively shorter. AA has the narrowest posteriors due to its high sensitivity.

\begin{figure}
\centering
 \includegraphics[width=\columnwidth, height=0.25\textheight, keepaspectratio]{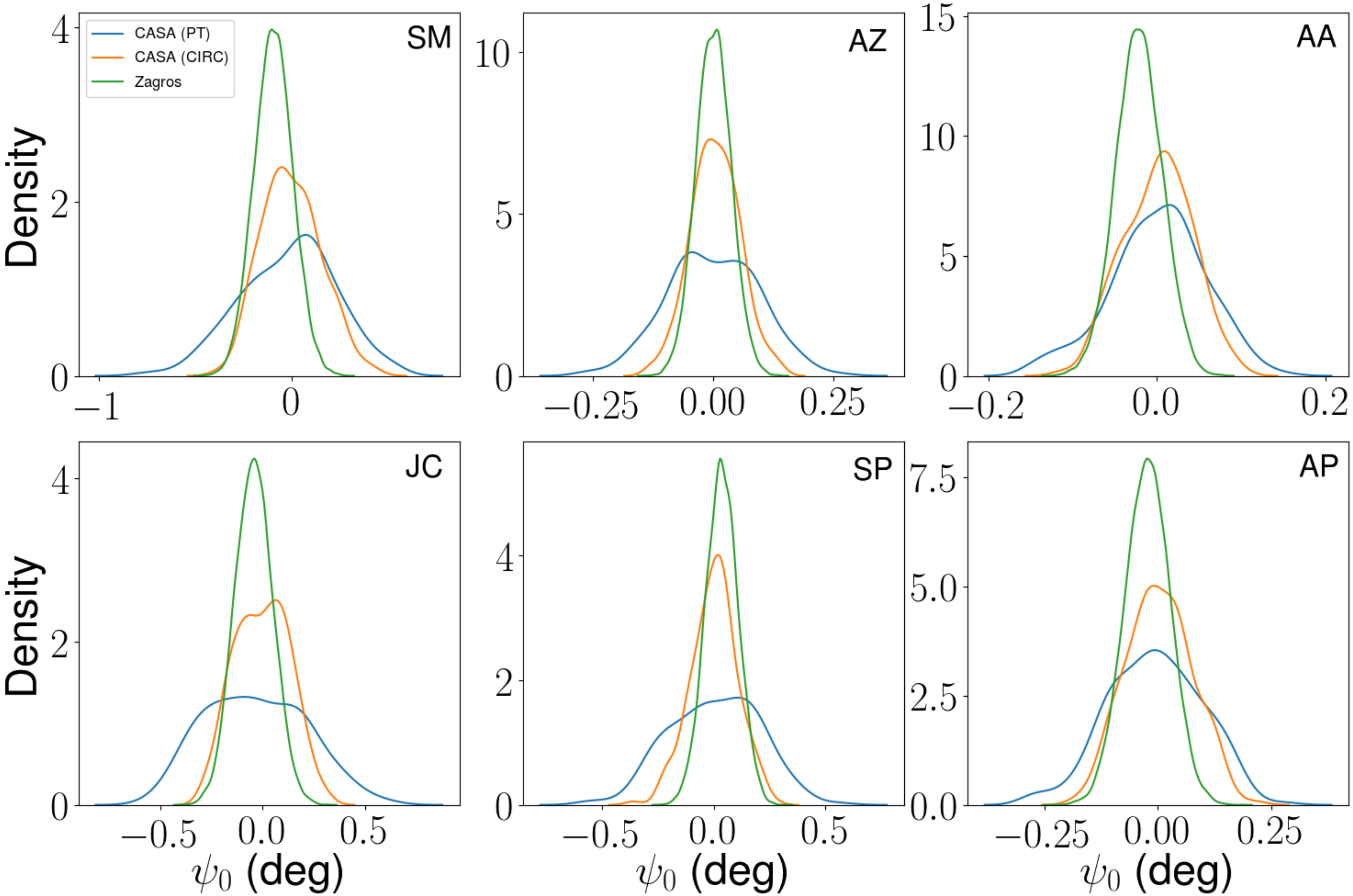}
 \caption[200 runs of casa fringe-fitting]{\textsc{zagros} posteriors (green) from Fig. \ref{fig:fwhm20uas_all_triplot} (20 $\upmu$as CIRC) shown alongside the histograms of the phase residuals obtained for 200 Monte Carlo simulations with different noise realisations using \textsc{casa} \texttt{fringefit}, with PT (blue) and CIRC (red) provided as input source models independently. The ground-truth values are shifted to zero.
 }
\label{fig:casavszagros_ph_CIRC}
\end{figure}

\begin{figure}
\centering
 \includegraphics[width=\columnwidth, height=0.25\textheight, keepaspectratio]{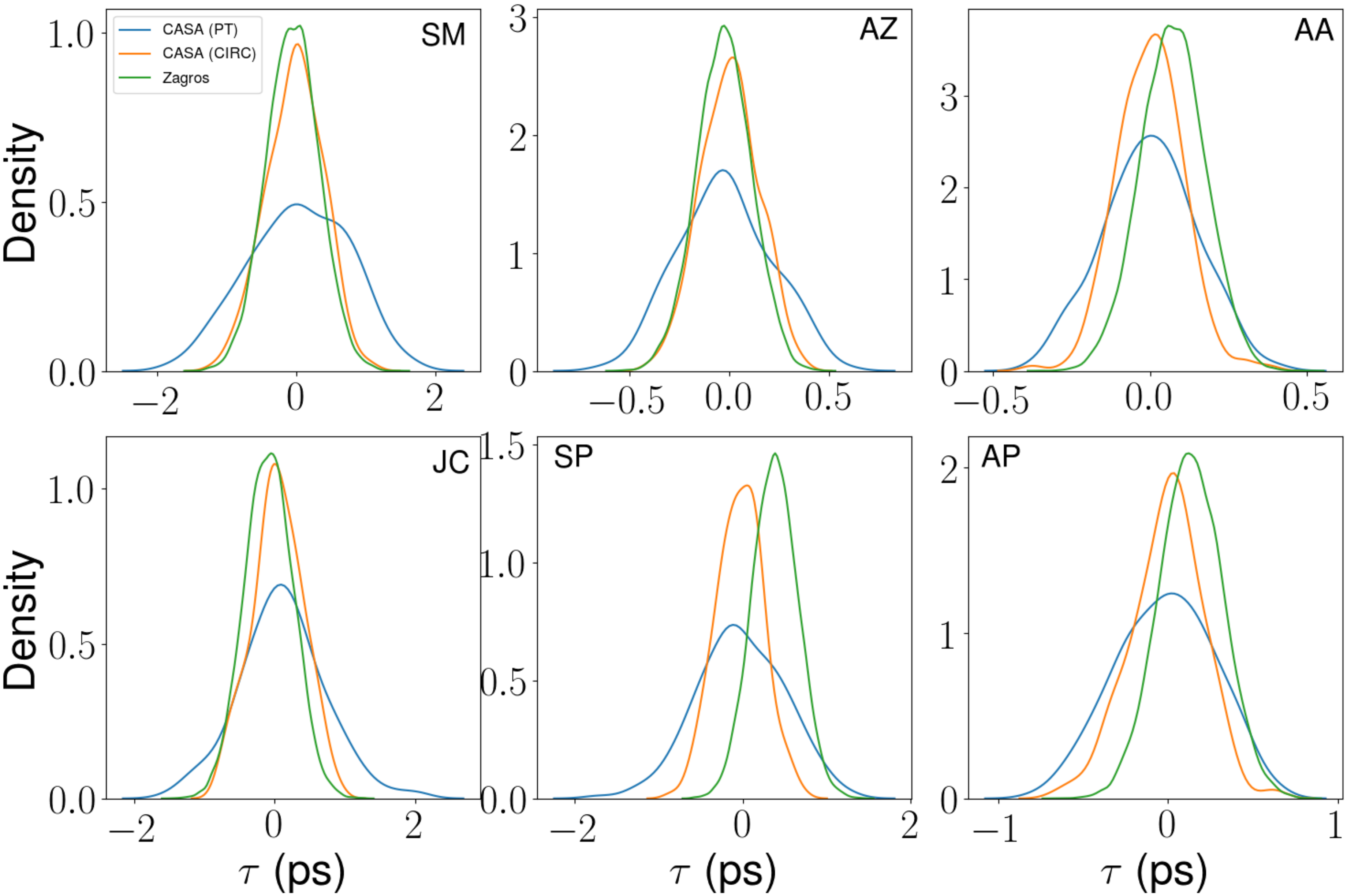}
 \caption[200 runs of casa fringe-fitting]{Same as Fig. \ref{fig:casavszagros_ph_CIRC}, but for delay residuals.
 }
\label{fig:casavszagros_CIRC}
\end{figure}

To obtain the distribution of \textsc{casa} \texttt{fringefit} results for comparison with \textsc{zagros}, we simulate 200 instances of each dataset with independent noise realisations, with the same rms $\sigma_{pq}$ in equation (\ref{eq:varyingsigma}). Following this, we fringe-fit each dataset twice, with and without incorporating the exact source model in the process, to obtain the best and worst possible estimates from \textsc{casa}.
Figs. \ref{fig:casavszagros_ph_CIRC} and \ref{fig:casavszagros_CIRC} show the results of this process.
Only those solutions with $\mathrm{SNR} \geq 3$ are included for global least-squares minimisation via the \texttt{minsnr} parameter. The phase residuals obtained with \texttt{fringefit} are re-referenced to the central frequency of the band. For this symmetric source structure, we see that the point source assumption during fringe-fitting is sufficient and the histograms of the \texttt{fringefit} results coincide well with the Bayesian posteriors.

\subsubsection{Elliptical Gaussian (ELLIP)}
\label{sss:ellip}
We next perform fringe-fitting on elliptical Gaussian models (ELLIP) of size 25 $\times$ 5 $\upmu$as oriented at a position angle of $60^{\circ}$. Model selection between PT and ELLIP on these data results in a Bayes factor on the order of $10^{12}$ (with the error in relative evidence being $\pm 0.66$), indicating a very high preference for the elliptical Gaussian. 

\begin{figure*}
\centering
 \includegraphics[scale=1.25]{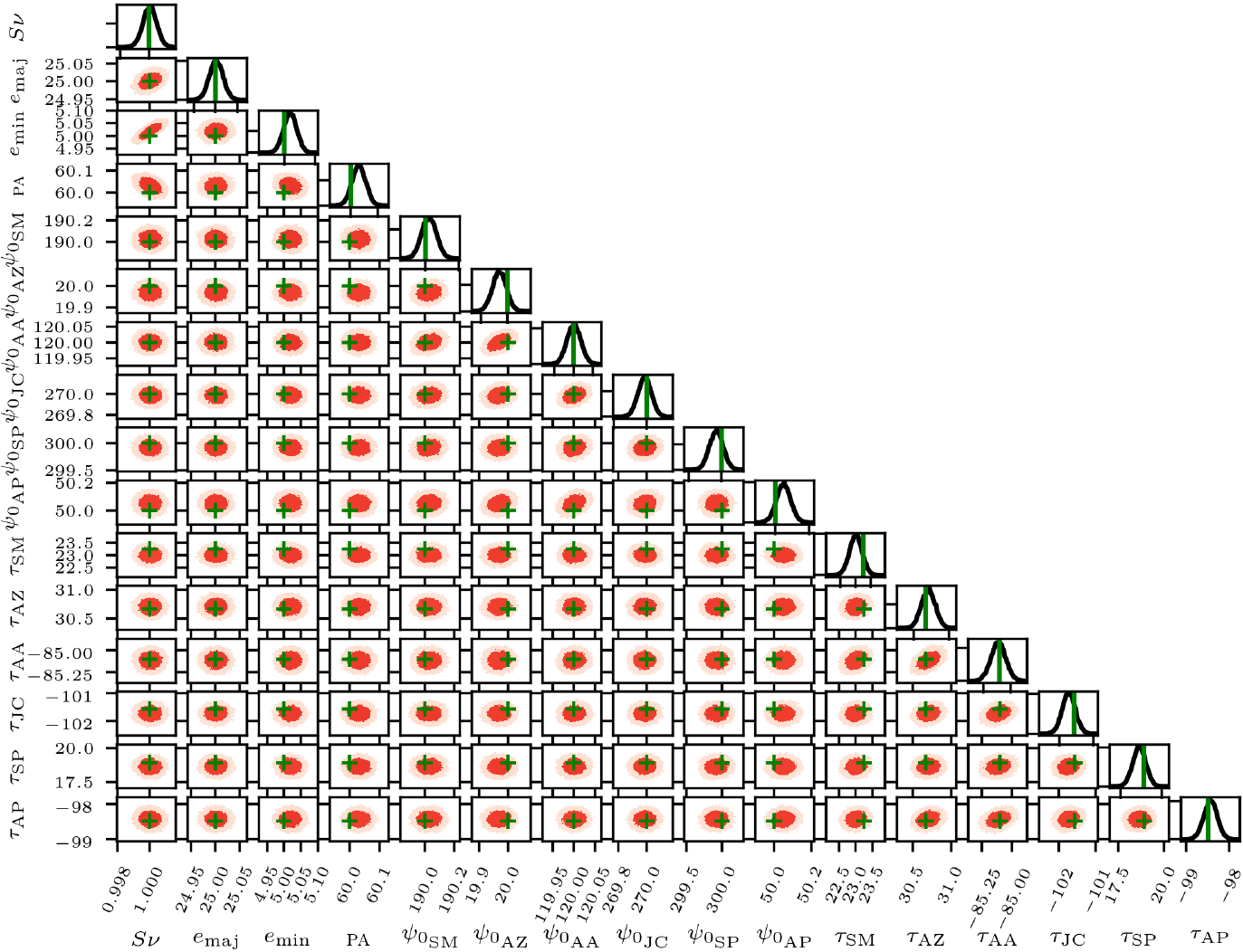}
 \caption[ELLIP bayesfringe results]{Same as Fig.~\ref{fig:fwhm20uas_all_triplot}, but for the model ELLIP of size 25 $\times$ 5 $\upmu$as.}
\label{fig:ellipgauss_delayoffsets_triplot}
\end{figure*}

The posterior distributions of the parameters corresponding to ELLIP are shown in Fig.~\ref{fig:ellipgauss_delayoffsets_triplot}.
Here too, we see that the Bayesian framework estimates the source parameters and the phase and delay residuals accurately, as shown by the vertical green lines indicating the true values. There are no significant degeneracies between the source and instrumental parameters, except for the slight correlation between the AA and AP delay estimates as with the CIRC model. There are no appreciable systematic offsets in the delay posteriors.

\begin{figure}
\centering
 \includegraphics[width=\columnwidth, height=0.25\textheight, keepaspectratio]{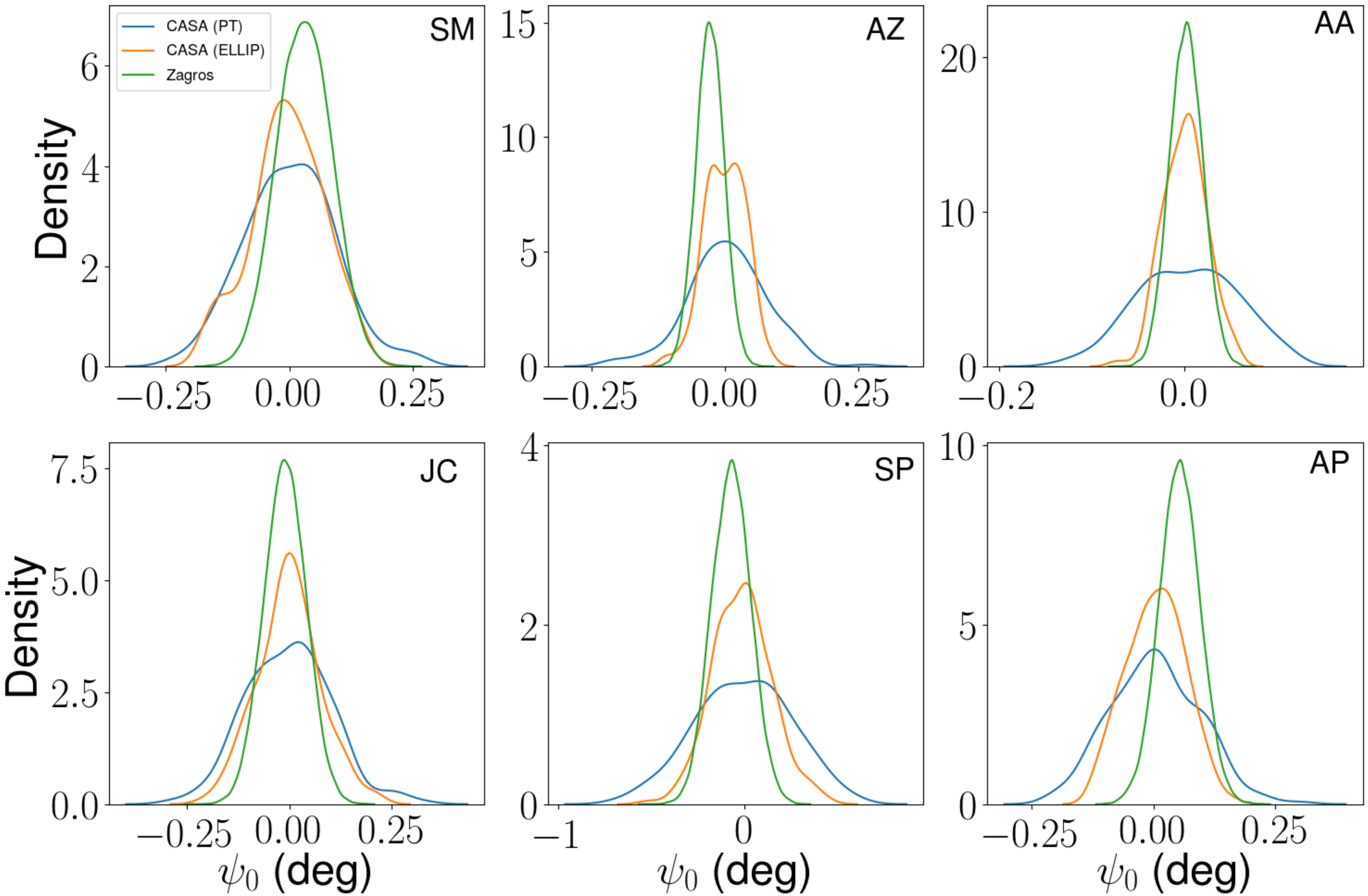}
 \caption[200 runs of casa fringe-fitting]{\textsc{zagros} posteriors (green) from Fig. \ref{fig:ellipgauss_delayoffsets_triplot} (25$\times$5 $\upmu$as ELLIP) shown alongside the histograms of the phase residuals obtained for 200 Monte Carlo simulations with different noise realisations using \textsc{casa} \texttt{fringefit}, with PT (blue) and ELLIP (red) provided as input source models independently. The ground-truth values are shifted to zero.}
\label{fig:casaellip_ph}
\end{figure}

\begin{figure}
\centering
 \includegraphics[width=\columnwidth, height=0.25\textheight, keepaspectratio]{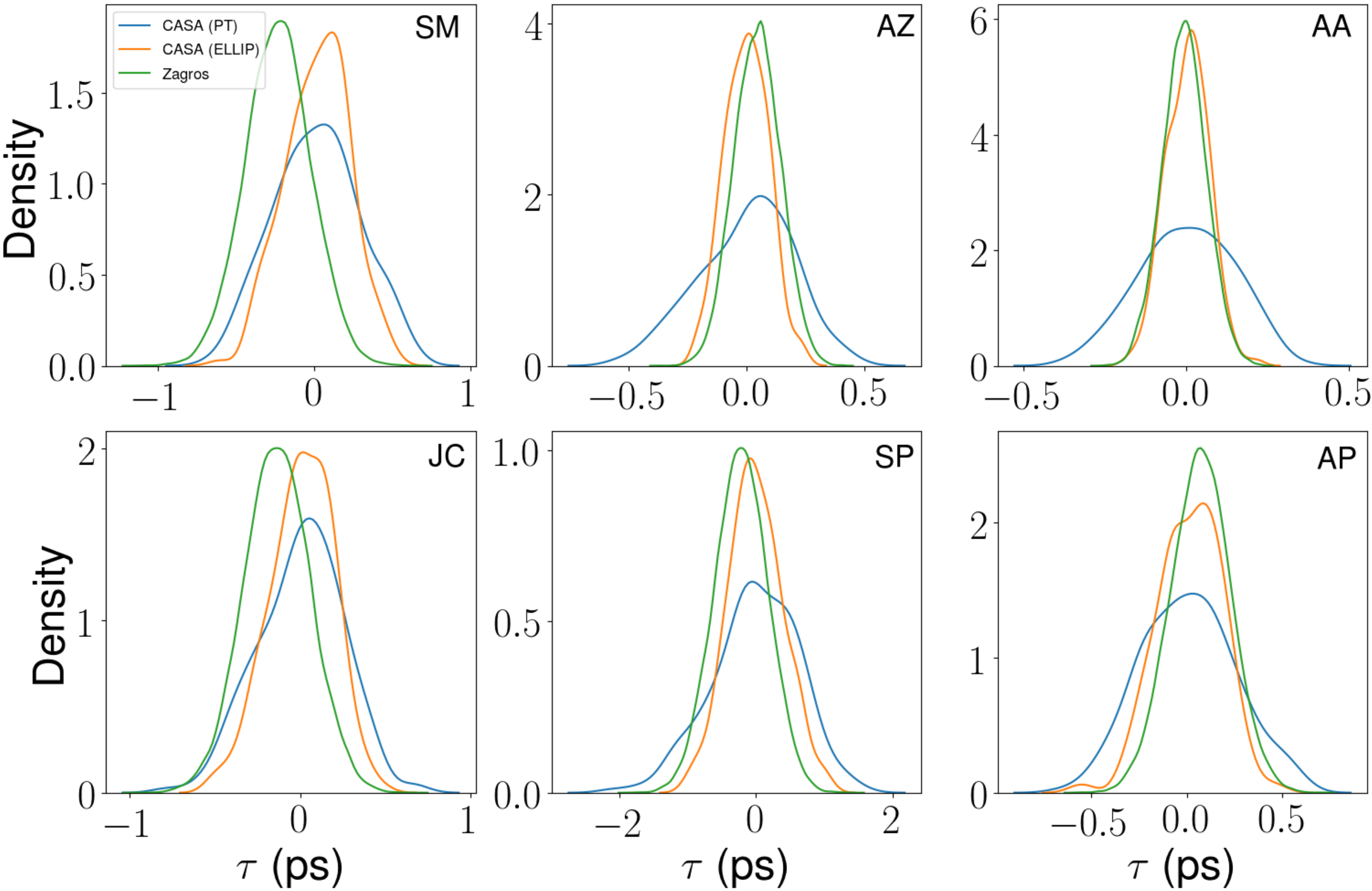}
 \caption[200 runs of casa fringe-fitting]{
 Same as Fig.~\ref{fig:casaellip_ph} but for delay residuals.}
\label{fig:casaellip}
\end{figure}

Figs.~\ref{fig:casaellip_ph} and \ref{fig:casaellip} respectively show the relative widths of the phase and delay estimates output by \textsc{casa} \texttt{fringefit} for 200 Monte Carlo simulations of the ELLIP source with different noise realisations.
As with CIRC, the phase and delay histograms coincide with the Bayesian posteriors. Neglecting the source structure for this centrally-located source distribution with two axes of symmetry does not affect the accuracy of the fringe-fitting.

\begin{figure}
\centering
 \includegraphics[width=\columnwidth, height=0.25\textheight, keepaspectratio]{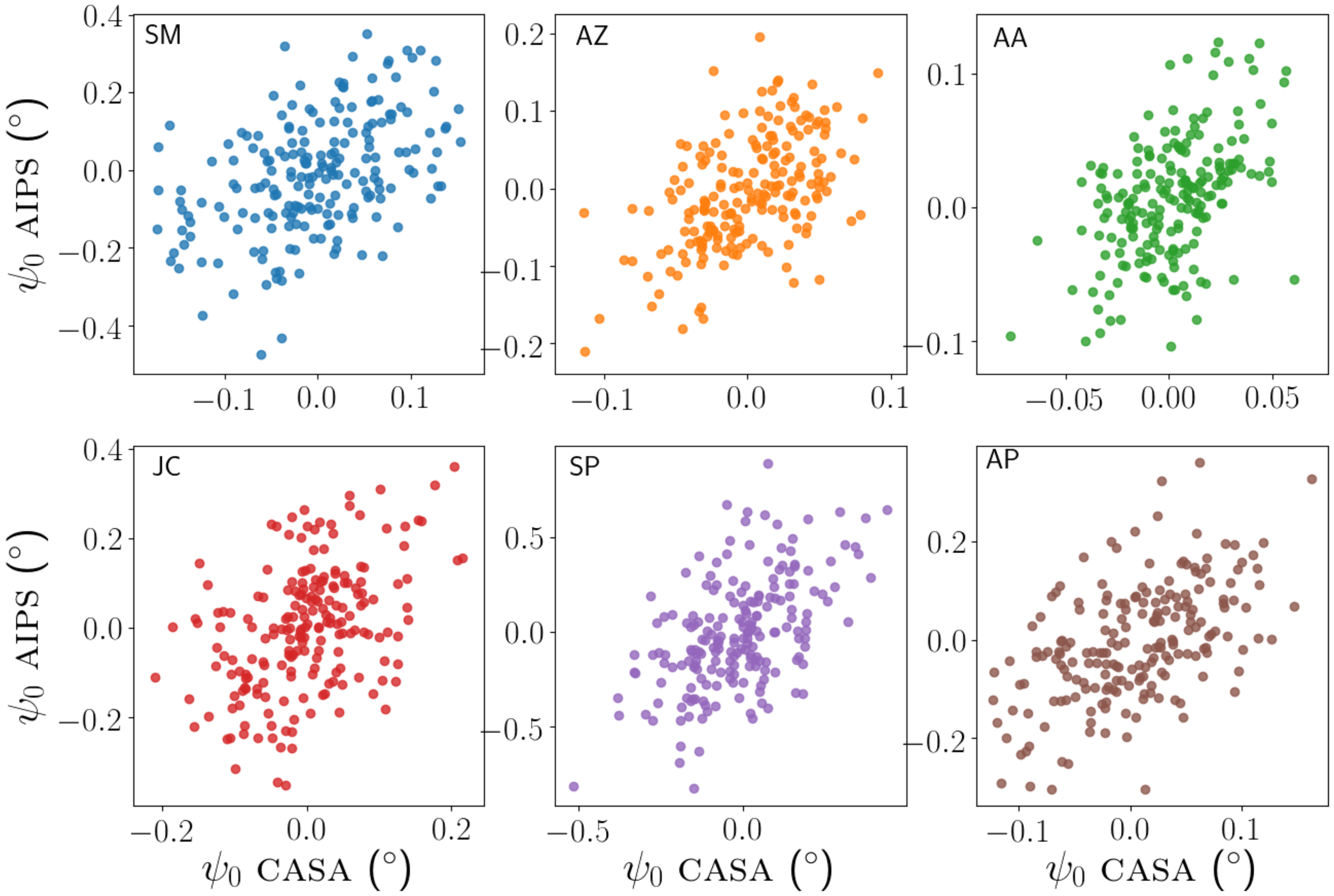}
 \caption[aips vs casa delays]{A comparison of \textsc{aips} and \textsc{casa} estimates of the phase residuals, with the true values shifted to zero, for the 200 synthetic datasets of the 25$\times$5 $\upmu$as ELLIP source with the exact source model input during fringe-fitting.}
\label{fig:aipscasaellipphases}
\end{figure}

\begin{figure}
\centering
 \includegraphics[width=\columnwidth, height=0.25\textheight, keepaspectratio]{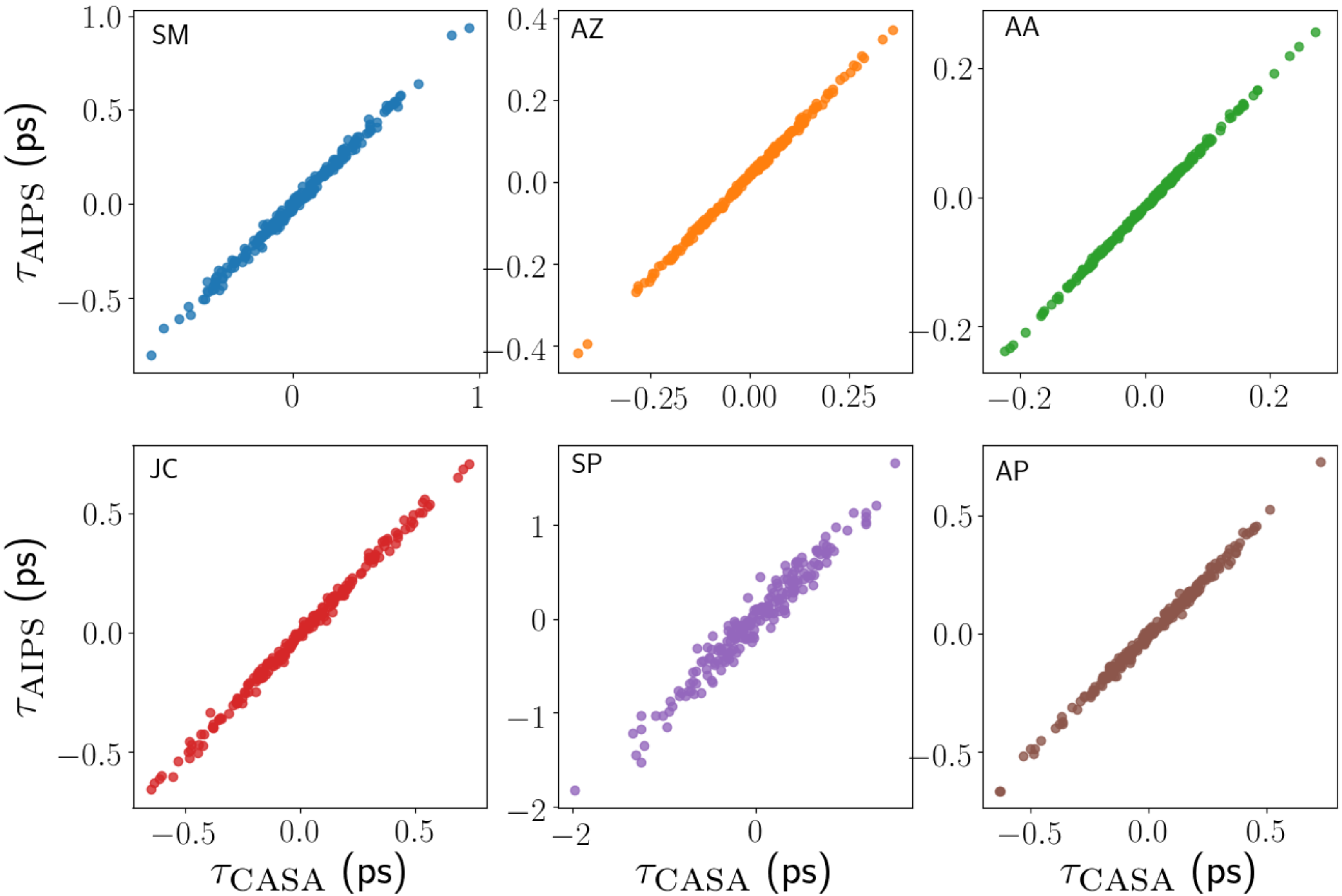}
 \caption[aips vs casa phases]{Same as Fig.~\ref{fig:aipscasaellipphases}, but for delay residuals.}
\label{fig:aipscasaellipdelays}
\end{figure}

For the ELLIP case, we also compare the \textsc{casa} estimates with the results of fringe-fitting this dataset using the \textsc{aips} task \texttt{FRING}. We choose this model for the \textsc{aips} comparison since the 2PT model which has more than one source cannot be input to the \texttt{FRING} task in \textsc{aips}. As with \texttt{fringefit}, the fringe-fitting is performed by inputting the exact source model to \texttt{FRING}. A comparison of the \texttt{fringefit} and \texttt{FRING} results is shown in Figs. \ref{fig:aipscasaellipphases} and \ref{fig:aipscasaellipdelays}).
The results from both \textsc{aips} and \textsc{casa} are re-referenced to the same frequency (the central frequency of the band) used by \textsc{zagros}. There is almost perfect correlation between the \textsc{aips} and \textsc{casa} delay residual estimates, while the phase residual estimates also follow the same trend, but are more loosely correlated. This correspondence between \textsc{casa}, \textsc{aips}, and \textsc{zagros} results indicates that our software is consistent with different implementations of the fringe-fitting algorithm.

\subsubsection{Two point sources (2PT)}
\label{sss:2pt}
The next class of source models we test, 2PT, are asymmetric source models with a primary point source of flux density 1 Jy at the phase centre and a secondary point source of 0.3 Jy located away from the central source at different distances. While the flux density ratios and source separations may change, the overall structure in this source model is more typical of VLBI sources and therefore is an important test to perform.

We simulate 9 datasets in which the secondary source is located at varying distances from the central source, from 20 $\upmu$as to 100 $\upmu$as in Declination ($\Delta \alpha = 0$), in steps of 10 $\upmu$as. The flux densities of both the sources and the location of the secondary source are allowed to vary with uniform priors with the hyperparameters shown in Table~\ref{tab:simpriors}. Crucially, the position prior of the secondary source is allowed to overlap with that of the primary source. In each case, the correct model (2PT) was favoured with a Bayes factor on the order of $10^{12}$. The error in relative evidence is $\pm 0.7$.

\begin{figure*}
\centering
 \includegraphics[scale=1.25]{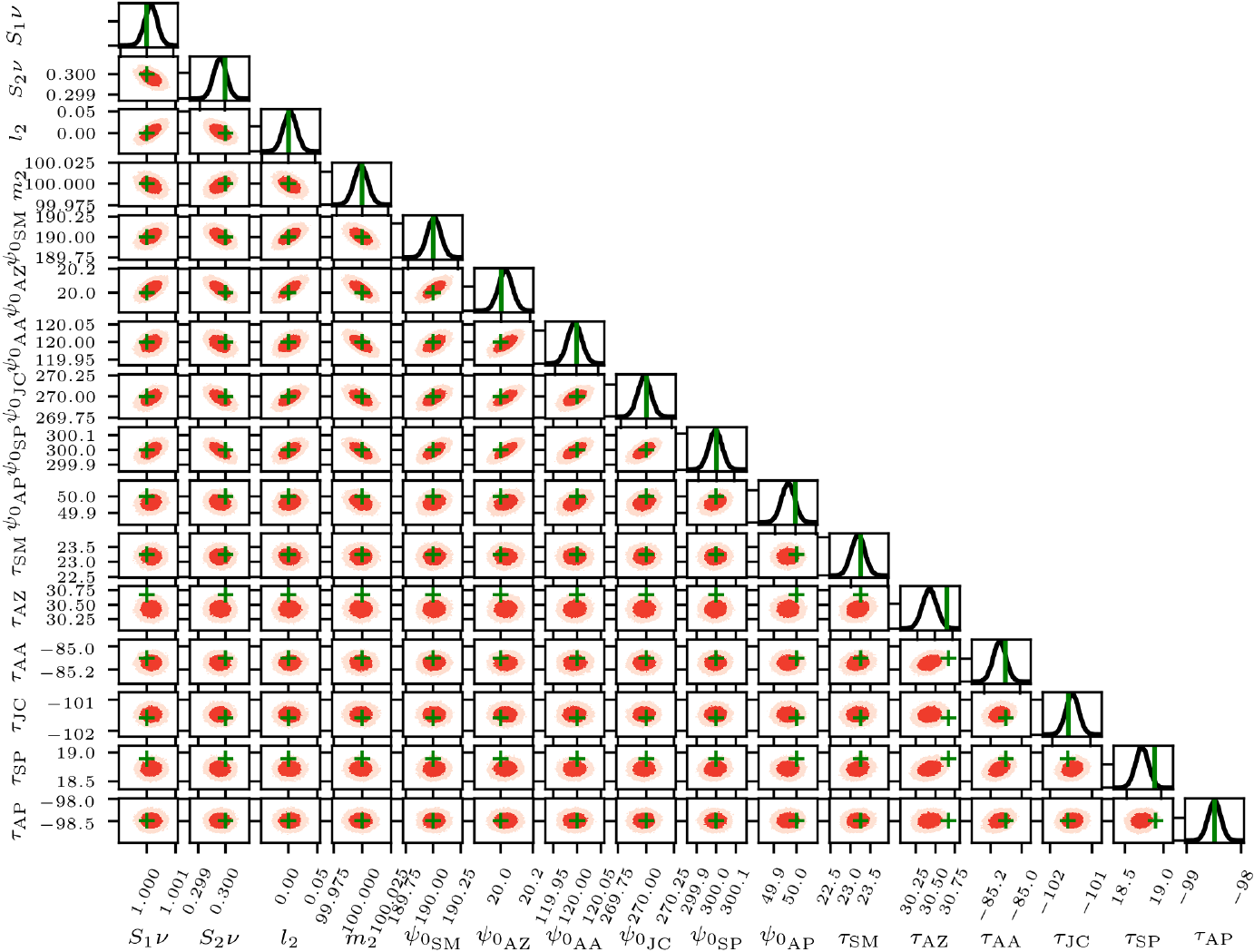}
 \caption[2PT bayesfringe results]{Same as Fig.~\ref{fig:fwhm20uas_all_triplot}, but for the model 2PT with the second source located at $\Delta \delta = 100\, \upmu$as.}
\label{fig:off100uas_triplot}
\end{figure*}

Fig.~\ref{fig:off100uas_triplot} shows the posterior distributions of the parameters of 2PT, when the secondary source is located 100 $\upmu$as away from the centre.
This asymmetric source structure reveals stronger correlations between the source structure and phase residuals. The phase residuals of the different stations are also correlated with each other. The delay residuals are independent of the source structure, as with the previous models. 

\begin{figure}
\centering
 \includegraphics[width=\columnwidth, height=0.25\textheight, keepaspectratio]{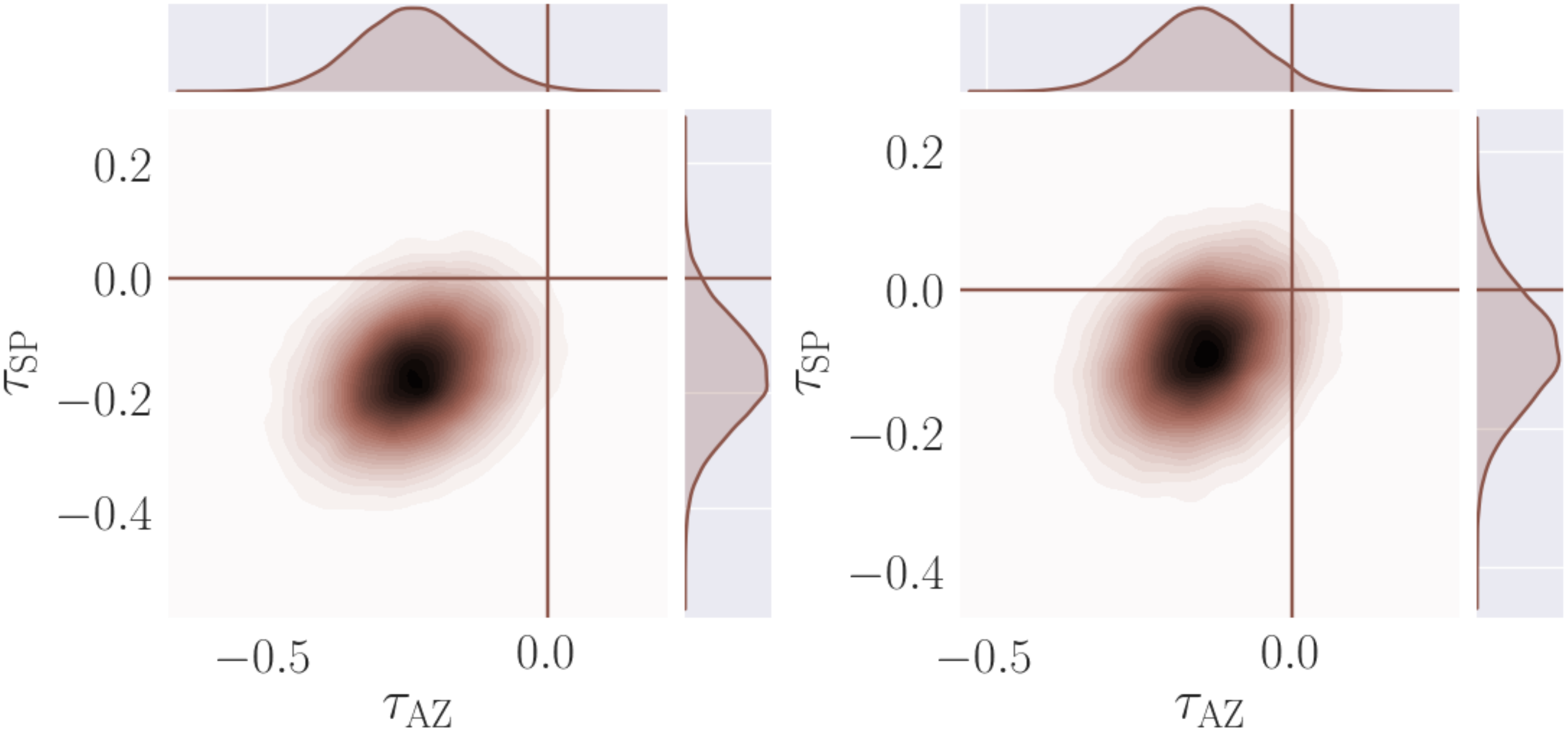}
 \caption[2PT compare-azsp]{Comparison of the delay posteriors of AZ and SP from Fig. \ref{fig:off100uas_triplot} (left) with those obtained for synthetic data generated with the same sky model rotated by $90^{\circ}$ ($\Delta \alpha=-100\, \upmu$as, $\Delta \delta=0\, \upmu$as).}
\label{fig:azsp2d}
\end{figure}

\begin{figure}
\centering
 \includegraphics[width=\columnwidth, height=0.25\textheight, keepaspectratio]{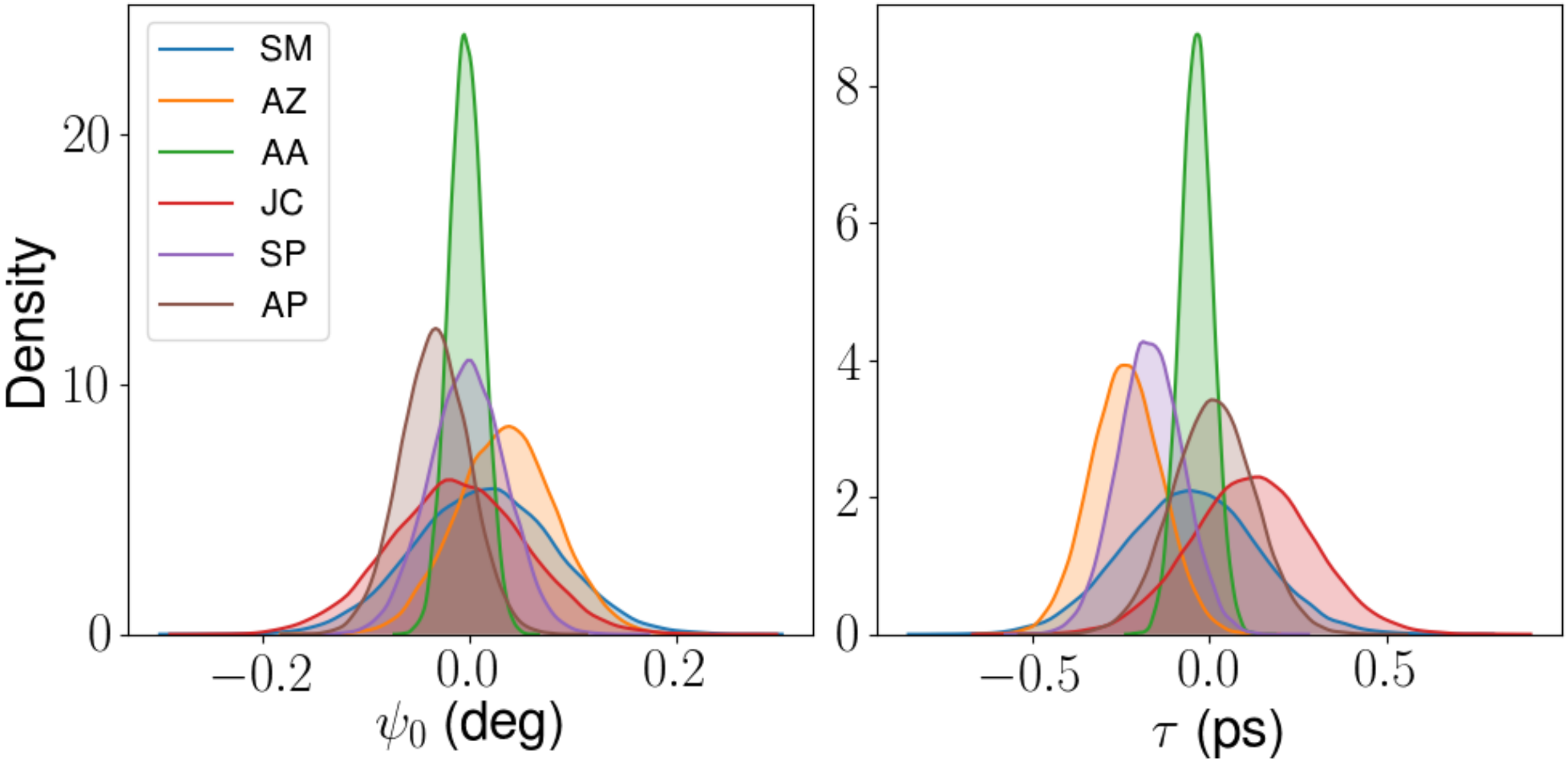}
 \caption[TWOPT-phdl-blsnr]{The posteriors of the phase and delay residuals from Fig. \ref{fig:off100uas_triplot} ($\Delta \alpha=0\, \upmu$as, $\Delta \delta=100\, \upmu$as) for all antennas. The ground-truth values are shifted to zero.}
\label{fig:TWOPT_phdl_blsnr}
\end{figure}

The ground-truth values for almost all fringe-fitting parameters are located within a credible interval of $1\sigma$, except for the delay residuals of AZ and SP. For these two antennas, the ground-truth values lie at the $2\sigma$ level of the posteriors. This could be due to contributions from various factors such as the source orientation with respect to the PSF of the array, which in turn is a function of the baseline orientation. To test this, we generate synthetic data with a 2PT source model rotated by $90^{\circ}$ in the sky ($\Delta \alpha = -100\, \upmu$as, $\Delta \delta = 0$) and perform fringe-fitting with \textsc{zagros}. Fig. \ref{fig:azsp2d} shows the comparison between the delay posteriors of AZ and SP obtained for this dataset and that shown in Fig. \ref{fig:off100uas_triplot}.
We see that the systematic offsets in the posterior peaks for both AZ and SP are reduced. Fig. \ref{fig:TWOPT_phdl_blsnr} also shows the relative widths and the magnitude of the spread of the posteriors from Fig. \ref{fig:off100uas_triplot} around the ground-truth values for all antennas plotted together. All the posteriors have comparable widths, with AA being the narrowest due to its high sensitivity.

\begin{figure}
\centering
 \includegraphics[width=\columnwidth, height=0.25\textheight, keepaspectratio]{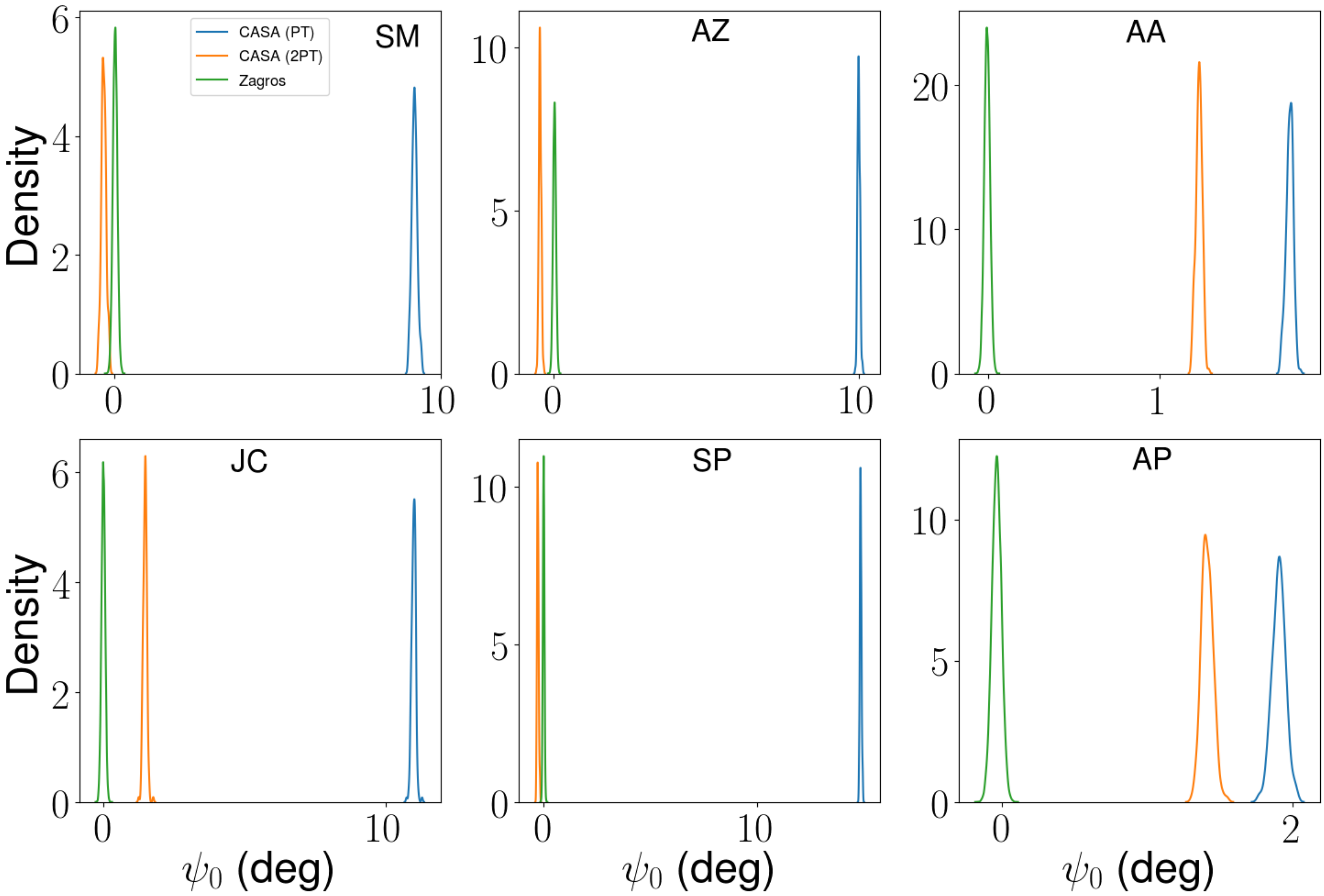}
 \caption[off100 casa delays]{\textsc{zagros} posteriors (green) from Fig. \ref{fig:off100uas_triplot} shown alongside the histograms of the phase residuals obtained for 200 Monte Carlo simulations with different noise realisations using \textsc{casa} \texttt{fringefit}, with PT (blue) and 2PT (red) provided as input source models independently. The ground-truth values are shifted to zero.}
\label{fig:casavsbff_ph_2PT}
\end{figure}

\begin{figure}
\centering
 \includegraphics[width=\columnwidth, height=0.25\textheight, keepaspectratio]{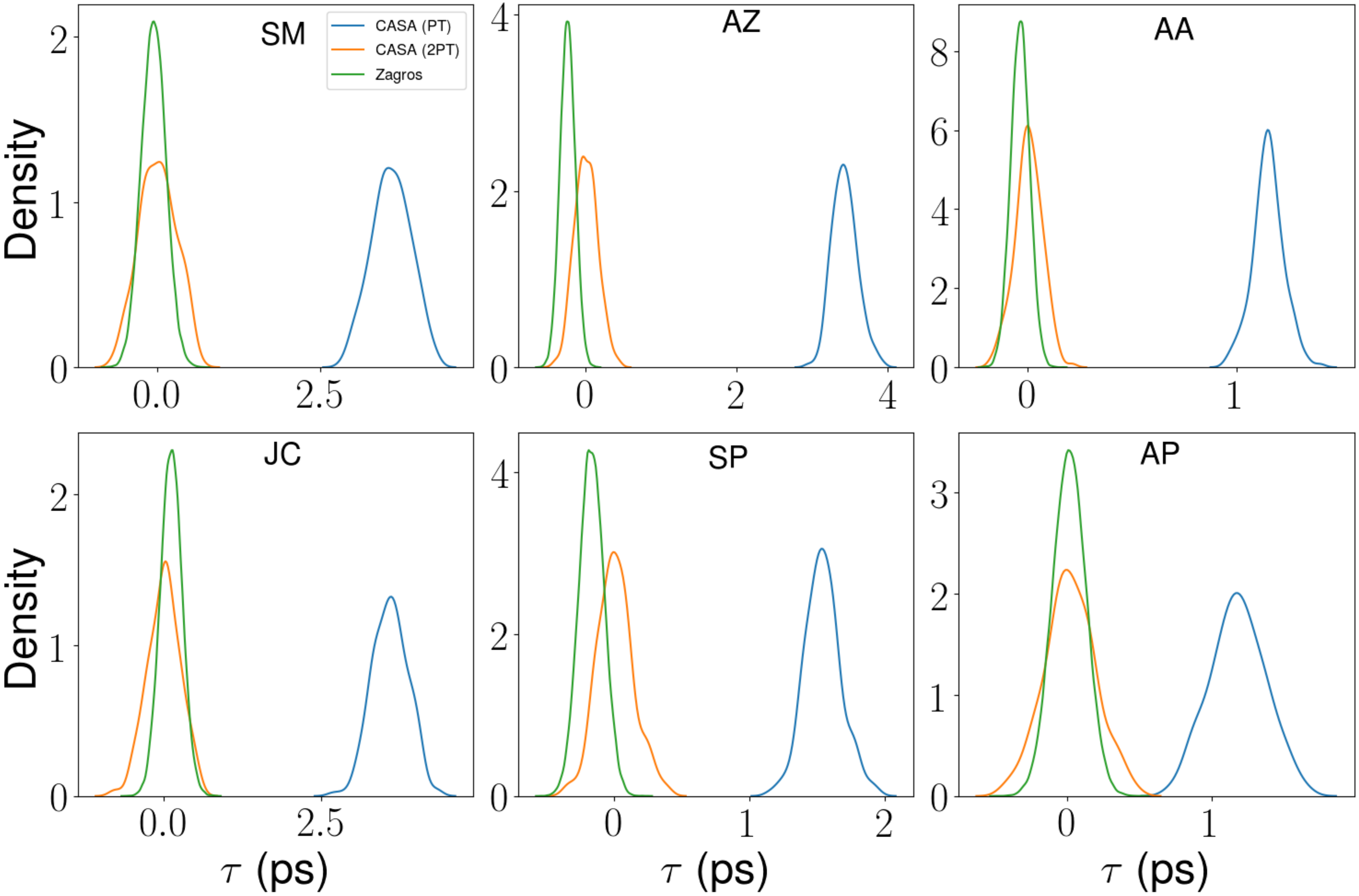}
 \caption[off100 casa delays]{Same as Fig. \ref{fig:casavsbff_ph_2PT}, but for delay residuals.}
\label{fig:casavsbff_2PT}
\end{figure}

\begin{figure}
\centering
 \includegraphics[width=\columnwidth, height=0.25\textheight, keepaspectratio]{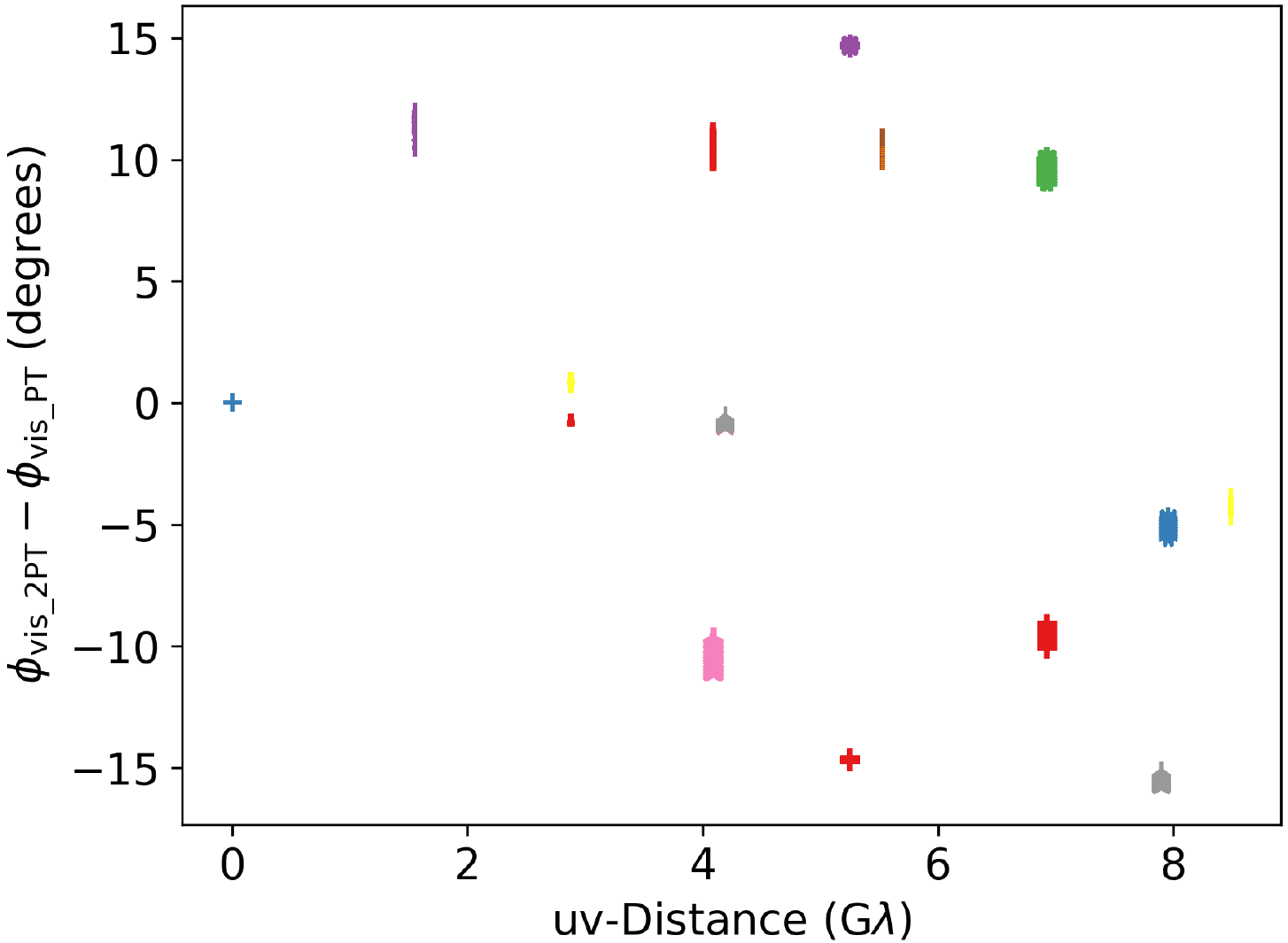}
 \caption[comparecasaphases]{Difference between the corrected phases for the 2PT simulation, after fringe-fitting with and without incorporating the source model, shown as a function of baseline length. The data are coloured by baseline and correspond to all channels and time.}
\label{fig:comparecasaphases}
\end{figure}

Fig.~\ref{fig:casavsbff_ph_2PT} compares the \textsc{zagros} posteriors with \textsc{casa} estimates.
Here, the point source assumption introduces significant offsets of up to $\pm$ 15$^{\circ}$ in the estimated phase residuals. Incorporating the exact source structure in \textsc{casa} results in the estimated phases coinciding with the \textsc{zagros} estimates. Fig.~\ref{fig:casavsbff_2PT} shows the comparison between the delay estimates obtained with both \textsc{zagros} and \texttt{fringefit}.
In this case, not accounting for the source structure results in offsets of up to $\pm 4$ ps in the estimated delay residuals. Figure \ref{fig:comparecasaphases} shows the differences in phases between the corrected visibilities, after fringe-fitting with and without incorporating the correct source model.
We may expect these offsets to be much larger in the low SNR regime. A crucial advantage of \textsc{zagros} is that it can capture the uncertainties in and degeneracies between the source parameters and the fringe-fitting related parameters completely, allowing one to draw more robust scientific inferences in the more difficult cases.

\section{Discussion}
\label{sec:disc}
As the source structure becomes more complex, it must be incorporated in the fringe-fitting process to avoid significant offsets in the corrected visibility phases. We find that our probabilistic approach to fringe-fitting performs at least as well as other widely-used software packages used for fringe-fitting, if not better. This is especially true in cases where a good source model from previous observations is unavailable or is difficult to obtain from the data, as is normally the case when the \emph{uv}-coverage is sparse. It also possesses the unique advantage of returning full posterior distributions for the model parameters and thus reliable propagation of uncertainties and the use of more informative priors further down in the calibration process where iterative calibration is necessary.

We find that for resolved Gaussian source morphologies at the phase centre, the effects of not incorporating the source model is not as significant as for an asymmetric source model such as two point sources. While for short-duration observations, the sparse sampling will ultimately constrain the amount of information that can be obtained on source structure, especially if the source structure is complex, the availability of posteriors on super-resolved source parameters provides a mechanism for establishing the veracity of the results obtained using conventional methods. For simple symmetric sources, the conventional methods generally yield results that are close to the truth which is borne out by the correspondence with the Bayesian posteriors. For asymmetric source distributions, any unmodelled structure not accounted for during fringe-fitting introduces systematic offsets in the estimates of the phase and delay residuals, as evident from the point source assumption using \textsc{casa} \texttt{fringefit} and the relative Bayesian evidences that are orders of magnitude (up to $10^{12}\! :\! 1$) in favour of the correct source model in \textsc{zagros}.

Furthermore, the unmodelled flux could be subsumed into self-calibration solutions, causing artefacts and suppressing real sources, even in the case of connected-element interferometers with a large number of stations \citep{trienko2014, sob2019}. VLBI observations can be especially affected by this, given the sparsity of \emph{uv}-coverage. Calibration and imaging is an ill-posed problem, with the solutions selected by self-calibration strongly affected by initial conditions. Obtaining joint posteriors of source parameters while performing fringe-fitting allows for the starting conditions to be closer to the ground-truth. In cases where the source structure is comparable to or smaller than the size of the PSF, it is difficult to obtain a model as a result of a previous imaging step. In such cases, using a priori knowledge of the source structure, modelled even as simple geometric models, selected based on Bayesian evidences obtained by a relatively quick run of \textsc{zagros}, provides a good starting point for calibration; this would be especially useful in the low SNR regime.

Application of this technique to synthetic data with low SNR will help set an upper limit to the effect that the systematic offsets of the posteriors from the ground-truth will have on subsequent calibration of actual VLBI data. The correlation between source parameters and phase and delay residuals may be relevant to specialist experiments mentioned in Section \ref{sec:intro}.

Apart from imaging experiments, this technique will find special applications in high precision astrometry and geodesy. For instance, simultaneous estimation of source positions in the presence of phase residuals on simulated European VLBI Network (EVN) data at 6.7 GHz, yield highly accurate estimates of the source positions \citep{huib2019}. Reversing this assumption and treating the source positions as known and the antenna locations as unknown, we can measure geodetic quantities more accurately \citep[][Chapter 23]{Syn1999}. The crucial advantage to geodetic experiments here is the capability to estimate the frequency-dependent core-shifts of sources simultaneously with the antenna positions and study the degeneracies between them, since it is straightforward to account for frequency-dependence in the source structure in our current framework.


\section{Computational Considerations and Outlook}
\label{sec:future}
The Bayesian analyses performed for this paper take a maximum of about 8 hours to complete on a machine with an NVIDIA Tesla K20m GPU, compared to the $\sim30$ minutes required for obtaining the histograms by fringe-fitting 200 datasets using \textsc{casa} \texttt{fringefit} (excluding the $\sim2$ hours required to simulate 200 datasets using \textsc{meqsilhouette}) on the same machine. In order to systematically explore a larger parameter space with \textsc{zagros} in a reasonable amount of time (e.g. more complex source models, time and frequency variable complex gains, rate residuals, etc.), faster model computation and sampling techniques become necessary.
 For larger visibility data set sizes, the forward-modelling step must be distributed over multiple CPU and/or GPU nodes due to increased memory requirements, which therefore necessitates processing on High Performance Computing (HPC) clusters. A distributed version of \textsc{codex-africanus} (Perkins et al., in prep.) and the corresponding \textsc{zagros} version which can distribute model computation between multiple HPC nodes using the \textsc{dask}\footnote{\url{https://dask.org}.} library are currently under development, and will be available for production in the near future. Including complex-valued, time and frequency varying delay and rate residuals is a straightforward extension of the current framework that would benefit from increased computational speed.

The geometric models explored in this paper have relatively simple structures, modelled using point sources and Gaussians. While the asymmetric source models with secondary point sources placed at different locations is typical of some VLBI sources and are useful for studying the limitations of traditional fringe-fitting, the primary EHT targets have much more complex structure \citep[e.g.][]{eht1, eht4}.
We plan to generate synthetic data with geometric ring and crescent models with jet components and fringe-fit them in two different ways: (i) using multiple Gaussian components to capture the source structure in the RIME and (ii) constructing the source-related terms of the RIME by performing Fast Fourier Transform (FFT) on parametrised ring and jet structures input in FITS format. Comparing the fringe-fitting results of the two methods will reveal the limitations of using simple geometric models to model rings and crescents.

Knowledge of these systematics will enable us to choose between these methods to fringe-fit synthetic observations of general relativistic magneto hydrodynamical (GRMHD) models \citep[e.g.][]{monika2016} of the photon ring surrounding the shadow of a black hole \citep[e.g.][]{eht1, eht5}. Using the FFT method, important physical parameters necessary to build GRMHD models can be directly sampled in image space and posteriors obtained at the end of a \textsc{zagros} run. A systematic study of these fringe-fitting approaches, followed by a suite of self-calibration procedures to measure differences in the final maps is a natural extension of this and will be the subject of a future paper.

\citet{blackburn2020} have shown that the usage of closure quantities is equivalent to numerical marginalisation over unconstrained gains. Under the RIME framework, scalar closure relations generalise to matrix closure relations and apply under certain assumptions of source symmetry \citep{oms2011}. Marginalising over uniform priors on instrumental phases and using these estimates to initialise gains for the full inference problem may serve as a good computational improvement and will be explored. We also plan to extend these analyses to multiple spectral windows so that this framework can be used to estimate multi-band delays. The ability to include both spatially and frequency-resolved source structure in fringe-fitting will prove useful especially when fractional bandwidths are large, as will be the case for arrays such as the next generation Very Large Array (ngVLA) \citep{carilli2015} and the next generation EHT (ngEHT) \citep{blackburn2019astro2020}.

\section*{Acknowledgements}
We thank Simon Perkins for helpful discussions on using \textsc{codex-africanus}; Edward Higson, Will Handley, and Mike Hobson for discussions on \textsc{polychord} and \textsc{dypolychord}. We thank Lindy Blackburn, Bill Cotton, and Landman Bester for useful discussions. We also thank the anonymous referee for their helpful comments.

IN and RPD are grateful for the support from the New Scientific Frontiers with Precision Radio Interferometry Fellowship awarded by the South African Radio Astronomy Observatory (SARAO), which is a facility of the National Research Foundation (NRF), an agency of the Department of Science and Technology (DST) of South Africa. OS is supported by the South African Research Chairs Initiative of the Department of Science and Technology and the National Research Foundation. The authors acknowledge funding from the European Union's Horizon 2020 research and innovation programme JUMPING JIVE (grant agreement No 730884), the National Research Foundation (grant No 107324), and the Dutch Science Organization, NWO (contract No 629.003.009). All computation was carried out on the servers located at the Rhodes University Centre for Radio Astronomy Techniques and Technologies (RATT).





\bibliographystyle{mnras}
\bibliography{references} 

\begin{thebibliography}{}
\makeatletter
\relax
\def\mn@urlcharsother{\let\do\@makeother \do\$\do\&\do\#\do\^\do\_\do\%\do\~}
\def\mn@doi{\begingroup\mn@urlcharsother \@ifnextchar [ {\mn@doi@}
  {\mn@doi@[]}}
\def\mn@doi@[#1]#2{\def\@tempa{#1}\ifx\@tempa\@empty \href
  {http://dx.doi.org/#2} {doi:#2}\else \href {http://dx.doi.org/#2} {#1}\fi
  \endgroup}
\def\mn@eprint#1#2{\mn@eprint@#1:#2::\@nil}
\def\mn@eprint@arXiv#1{\href {http://arxiv.org/abs/#1} {{\tt arXiv:#1}}}
\def\mn@eprint@dblp#1{\href {http://dblp.uni-trier.de/rec/bibtex/#1.xml}
  {dblp:#1}}
\def\mn@eprint@#1:#2:#3:#4\@nil{\def\@tempa {#1}\def\@tempb {#2}\def\@tempc
  {#3}\ifx \@tempc \@empty \let \@tempc \@tempb \let \@tempb \@tempa \fi \ifx
  \@tempb \@empty \def\@tempb {arXiv}\fi \@ifundefined
  {mn@eprint@\@tempb}{\@tempb:\@tempc}{\expandafter \expandafter \csname
  mn@eprint@\@tempb\endcsname \expandafter{\@tempc}}}

\bibitem[\protect\citeauthoryear{{Alef} \& {Porcas}}{{Alef} \&
  {Porcas}}{1986}]{alefporcas1986}
{Alef} W.,  {Porcas} R.~W.,  1986, \aap, \href
  {http://adsabs.harvard.edu/abs/1986A%26A...168..365A} {168, 365}

\bibitem[\protect\citeauthoryear{{Blackburn} et~al.,}{{Blackburn}
  et~al.}{2019a}]{blackburn2019astro2020}
{Blackburn} L.,  et~al., 2019a, arXiv e-prints, \href
  {https://ui.adsabs.harvard.edu/abs/2019arXiv190901411B} {p. arXiv:1909.01411}

\bibitem[\protect\citeauthoryear{{Blackburn} et~al.,}{{Blackburn}
  et~al.}{2019b}]{blackburn2019}
{Blackburn} L.,  et~al., 2019b, \mn@doi [\apj] {10.3847/1538-4357/ab328d},
  \href {https://ui.adsabs.harvard.edu/abs/2019ApJ...882...23B} {882, 23}

\bibitem[\protect\citeauthoryear{{Blackburn}, {Pesce}, {Johnson}, {Wielgus},
  {Chael}, {Christian}  \& {Doeleman}}{{Blackburn}
  et~al.}{2020}]{blackburn2020}
{Blackburn} L.,  {Pesce} D.~W.,  {Johnson} M.~D.,  {Wielgus} M.,  {Chael}
  A.~A.,  {Christian} P.,   {Doeleman} S.~S.,  2020, \mn@doi [\apj]
  {10.3847/1538-4357/ab8469}, \href
  {https://ui.adsabs.harvard.edu/abs/2020ApJ...894...31B} {894, 31}

\bibitem[\protect\citeauthoryear{{Blecher}, {Deane}, {Bernardi}  \&
  {Smirnov}}{{Blecher} et~al.}{2017}]{tariq2017}
{Blecher} T.,  {Deane} R.,  {Bernardi} G.,   {Smirnov} O.,  2017, \mn@doi
  [Monthly Notices of the Royal Astronomical Society] {10.1093/mnras/stw2311},
  \href {http://adsabs.harvard.edu/abs/2017MNRAS.464..143B} {464, 143}

\bibitem[\protect\citeauthoryear{{Broderick} \& {Loeb}}{{Broderick} \&
  {Loeb}}{2009}]{broderick2009}
{Broderick} A.~E.,  {Loeb} A.,  2009, \mn@doi [The Astrophysical Journal]
  {10.1088/0004-637X/697/2/1164}, \href
  {http://adsabs.harvard.edu/abs/2009ApJ...697.1164B} {697, 1164}

\bibitem[\protect\citeauthoryear{Cappallo}{Cappallo}{2017}]{fourfit}
Cappallo R.,  2017, FOURFIT user's manual.
MIT Haystack Observatory

\bibitem[\protect\citeauthoryear{Carilli \& Holdaway}{Carilli \&
  Holdaway}{1999}]{carilli1999}
Carilli C.~L.,  Holdaway M.~A.,  1999, \mn@doi [Radio Science]
  {10.1029/1999RS900048}, 34, 817

\bibitem[\protect\citeauthoryear{{Carilli} et~al.,}{{Carilli}
  et~al.}{2015}]{carilli2015}
{Carilli} C.~L.,  et~al., 2015, arXiv e-prints, \href
  {https://ui.adsabs.harvard.edu/abs/2015arXiv151006438C} {p. arXiv:1510.06438}

\bibitem[\protect\citeauthoryear{{Cornwell} \& {Wilkinson}}{{Cornwell} \&
  {Wilkinson}}{1981}]{cornwellwilk1981}
{Cornwell} T.~J.,  {Wilkinson} P.~N.,  1981, \mn@doi [Monthly Notices of the
  Royal Astronomical Society] {10.1093/mnras/196.4.1067}, \href
  {http://adsabs.harvard.edu/abs/1981MNRAS.196.1067C} {196, 1067}

\bibitem[\protect\citeauthoryear{{Cotton}}{{Cotton}}{1995}]{cotton1995}
{Cotton} W.~D.,  1995, in {Zensus} J.~A.,  {Diamond} P.~J.,   {Napier} P.~J.,
  eds,  Astronomical Society of the Pacific Conference Series Vol. 82, Very
  Long Baseline Interferometry and the VLBA. p.~189

\bibitem[\protect\citeauthoryear{{En{\ss}lin}}{{En{\ss}lin}}{2018}]{ensslin2018}
{En{\ss}lin} T.~A.,  2018, arXiv e-prints, \href
  {https://ui.adsabs.harvard.edu/\#abs/2018arXiv180403350E} {p.
  arXiv:1804.03350}

\bibitem[\protect\citeauthoryear{{En{\ss}lin}, {Frommert}  \&
  {Kitaura}}{{En{\ss}lin} et~al.}{2009}]{ensslin2009}
{En{\ss}lin} T.~A.,  {Frommert} M.,   {Kitaura} F.~S.,  2009, \mn@doi [\prd]
  {10.1103/PhysRevD.80.105005}, \href
  {https://ui.adsabs.harvard.edu/\#abs/2009PhRvD..80j5005E} {80, 105005}

\bibitem[\protect\citeauthoryear{{En{\ss}lin}, {Junklewitz}, {Winderling},
  {Greiner}  \& {Selig}}{{En{\ss}lin} et~al.}{2014}]{ensslin2014}
{En{\ss}lin} T.~A.,  {Junklewitz} H.,  {Winderling} L.,  {Greiner} M.,
  {Selig} M.,  2014, \mn@doi [Physical Review E] {10.1103/PhysRevE.90.043301},
  \href {http://adsabs.harvard.edu/abs/2014PhRvE..90d3301E} {90, 043301}

\bibitem[\protect\citeauthoryear{{Event Horizon Telescope Collaboration}
  et~al.,}{{Event Horizon Telescope Collaboration} et~al.}{2019a}]{eht1}
{Event Horizon Telescope Collaboration} et~al., 2019a, \mn@doi [\apjl]
  {10.3847/2041-8213/ab0ec7}, \href
  {http://adsabs.harvard.edu/abs/2019ApJ...875L...1E} {875, L1}

\bibitem[\protect\citeauthoryear{{Event Horizon Telescope Collaboration}
  et~al.,}{{Event Horizon Telescope Collaboration} et~al.}{2019b}]{eht2}
{Event Horizon Telescope Collaboration} et~al., 2019b, \mn@doi [\apjl]
  {10.3847/2041-8213/ab0c96}, \href
  {http://adsabs.harvard.edu/abs/2019ApJ...875L...2E} {875, L2}

\bibitem[\protect\citeauthoryear{{Event Horizon Telescope Collaboration}
  et~al.,}{{Event Horizon Telescope Collaboration} et~al.}{2019c}]{eht3}
{Event Horizon Telescope Collaboration} et~al., 2019c, \mn@doi [\apjl]
  {10.3847/2041-8213/ab0c57}, \href
  {http://adsabs.harvard.edu/abs/2019ApJ...875L...3E} {875, L3}

\bibitem[\protect\citeauthoryear{{Event Horizon Telescope Collaboration}
  et~al.,}{{Event Horizon Telescope Collaboration} et~al.}{2019d}]{eht4}
{Event Horizon Telescope Collaboration} et~al., 2019d, \mn@doi [\apjl]
  {10.3847/2041-8213/ab0e85}, \href
  {http://adsabs.harvard.edu/abs/2019ApJ...875L...4E} {875, L4}

\bibitem[\protect\citeauthoryear{{Event Horizon Telescope Collaboration}
  et~al.,}{{Event Horizon Telescope Collaboration} et~al.}{2019e}]{eht5}
{Event Horizon Telescope Collaboration} et~al., 2019e, \mn@doi [\apjl]
  {10.3847/2041-8213/ab0f43}, \href
  {http://adsabs.harvard.edu/abs/2019ApJ...875L...5E} {875, L5}

\bibitem[\protect\citeauthoryear{{Event Horizon Telescope Collaboration}
  et~al.,}{{Event Horizon Telescope Collaboration} et~al.}{2019f}]{eht6}
{Event Horizon Telescope Collaboration} et~al., 2019f, \mn@doi [\apjl]
  {10.3847/2041-8213/ab1141}, \href
  {http://adsabs.harvard.edu/abs/2019ApJ...875L...6E} {875, L6}

\bibitem[\protect\citeauthoryear{{Falcke} \& {Markoff}}{{Falcke} \&
  {Markoff}}{2013}]{falcke2013}
{Falcke} H.,  {Markoff} S.~B.,  2013, \mn@doi [Classical and Quantum Gravity]
  {10.1088/0264-9381/30/24/244003}, \href
  {http://adsabs.harvard.edu/abs/2013CQGra..30x4003F} {30, 244003}

\bibitem[\protect\citeauthoryear{{Grobler}, {Nunhokee}, {Smirnov}, {van Zyl}
  \& {de Bruyn}}{{Grobler} et~al.}{2014}]{trienko2014}
{Grobler} T.~L.,  {Nunhokee} C.~D.,  {Smirnov} O.~M.,  {van Zyl} A.~J.,   {de
  Bruyn} A.~G.,  2014, \mn@doi [\mnras] {10.1093/mnras/stu268}, 439, 4030

\bibitem[\protect\citeauthoryear{Hamaker, Bregman  \& Sault}{Hamaker
  et~al.}{1996}]{hbs1996}
Hamaker J.~P.,  Bregman J.~D.,   Sault R.~J.,  1996, Astronomy \& Astrophysics,
  117, 137

\bibitem[\protect\citeauthoryear{{Handley}, {Hobson}  \& {Lasenby}}{{Handley}
  et~al.}{2015a}]{polychord1}
{Handley} W.~J.,  {Hobson} M.~P.,   {Lasenby} A.~N.,  2015a, \mn@doi [Monthly
  Notices of the Royal Astronomical Society] {10.1093/mnrasl/slv047}, \href
  {http://adsabs.harvard.edu/abs/2015MNRAS.450L..61H} {450, L61}

\bibitem[\protect\citeauthoryear{{Handley}, {Hobson}  \& {Lasenby}}{{Handley}
  et~al.}{2015b}]{polychord2}
{Handley} W.~J.,  {Hobson} M.~P.,   {Lasenby} A.~N.,  2015b, \mn@doi [Monthly
  Notices of the Royal Astronomical Society] {10.1093/mnras/stv1911}, \href
  {http://adsabs.harvard.edu/abs/2015MNRAS.453.4384H} {453, 4384}

\bibitem[\protect\citeauthoryear{Higson, Handley, Hobson  \& Lasenby}{Higson
  et~al.}{2019}]{Higson2019}
Higson E.,  Handley W.,  Hobson M.,   Lasenby A.,  2019, \mn@doi [Statistics
  and Computing] {10.1007/s11222-018-9844-0}, 29, 891

\bibitem[\protect\citeauthoryear{{Issaoun} et~al.,}{{Issaoun}
  et~al.}{2019}]{issaoun2019}
{Issaoun} S.,  et~al., 2019, \mn@doi [\apj] {10.3847/1538-4357/aaf732}, \href
  {https://ui.adsabs.harvard.edu/abs/2019ApJ...871...30I} {871, 30}

\bibitem[\protect\citeauthoryear{{Janssen} et~al.,}{{Janssen}
  et~al.}{2019}]{janssen2019}
{Janssen} M.,  et~al., 2019, \mn@doi [\aap] {10.1051/0004-6361/201935181},
  \href {https://ui.adsabs.harvard.edu/abs/2019A&A...626A..75J} {626, A75}

\bibitem[\protect\citeauthoryear{Jeffreys}{Jeffreys}{1961}]{jeffreys1961}
Jeffreys H.,  1961, Theory of Probability, 3rd edn.
Clarendon Press

\bibitem[\protect\citeauthoryear{Jones}{Jones}{1941}]{jones1941}
Jones R.~C.,  1941, Journal of the Optical Society of America

\bibitem[\protect\citeauthoryear{{Junklewitz}, {Bell}, {Selig}  \&
  {En{\ss}lin}}{{Junklewitz} et~al.}{2016}]{resolve2016}
{Junklewitz} H.,  {Bell} M.~R.,  {Selig} M.,   {En{\ss}lin} T.~A.,  2016,
  \mn@doi [Astronomy \& Astrophysics] {10.1051/0004-6361/201323094}, \href
  {http://adsabs.harvard.edu/abs/2016A%26A...586A..76J} {586, A76}

\bibitem[\protect\citeauthoryear{Kass \& Raftery}{Kass \&
  Raftery}{1995}]{kassraftery1995}
Kass R.~E.,  Raftery A.~E.,  1995, Journal of the American Statistical
  Association, 90, 773

\bibitem[\protect\citeauthoryear{Lochner, Natarajan, Zwart, Smirnov, Bassett,
  Oozeer  \& Kunz}{Lochner et~al.}{2015}]{biro2015}
Lochner M.,  Natarajan I.,  Zwart J.,  Smirnov O.,  Bassett B.,  Oozeer N.,
  Kunz M.,  2015, Monthly Notices of the Royal Astronomical Society, 450, 1308

\bibitem[\protect\citeauthoryear{{Mart{\'{\i}}-Vidal}, {P{\'e}rez-Torres}  \&
  {Lobanov}}{{Mart{\'{\i}}-Vidal} et~al.}{2012}]{martividal2012}
{Mart{\'{\i}}-Vidal} I.,  {P{\'e}rez-Torres} M.~A.,   {Lobanov} A.~P.,  2012,
  Astronomy \& Astrophysics, 541, A135

\bibitem[\protect\citeauthoryear{{Mo{\'s}cibrodzka}, {Falcke}  \&
  {Shiokawa}}{{Mo{\'s}cibrodzka} et~al.}{2016}]{monika2016}
{Mo{\'s}cibrodzka} M.,  {Falcke} H.,   {Shiokawa} H.,  2016, \mn@doi [\aap]
  {10.1051/0004-6361/201526630}, \href
  {https://ui.adsabs.harvard.edu/\#abs/2016A&A...586A..38M} {586, A38}

\bibitem[\protect\citeauthoryear{{Natarajan}, {Paragi}, {Zwart}, {Perkins},
  {Smirnov}  \& {van der Heyden}}{{Natarajan} et~al.}{2017}]{iniyan2017}
{Natarajan} I.,  {Paragi} Z.,  {Zwart} J.,  {Perkins} S.,  {Smirnov} O.,   {van
  der Heyden} K.,  2017, \mn@doi [Monthly Notices of the Royal Astronomical
  Society] {10.1093/mnras/stw2653}, \href
  {http://adsabs.harvard.edu/abs/2017MNRAS.464.4306N} {464, 4306}

\bibitem[\protect\citeauthoryear{{Readhead} \& {Wilkinson}}{{Readhead} \&
  {Wilkinson}}{1978}]{selfcal1978}
{Readhead} A.~C.~S.,  {Wilkinson} P.~N.,  1978, \mn@doi [The Astrophysical
  Journal] {10.1086/156232}, 223, 25

\bibitem[\protect\citeauthoryear{{Readhead}, {Walker}, {Pearson}  \&
  {Cohen}}{{Readhead} et~al.}{1980}]{readhead1980}
{Readhead} A.~C.~S.,  {Walker} R.~C.,  {Pearson} T.~J.,   {Cohen} M.~H.,  1980,
  \mn@doi [Nature] {10.1038/285137a0}, \href
  {http://adsabs.harvard.edu/abs/1980Natur.285..137R} {285, 137}

\bibitem[\protect\citeauthoryear{{Schwab} \& {Cotton}}{{Schwab} \&
  {Cotton}}{1983}]{schwabcotton1983}
{Schwab} F.~R.,  {Cotton} W.~D.,  1983, \mn@doi [The Astronomical Journal]
  {10.1086/113360}, \href {http://adsabs.harvard.edu/abs/1983AJ.....88..688S}
  {88, 688}

\bibitem[\protect\citeauthoryear{Smirnov}{Smirnov}{2011}]{oms2011}
Smirnov O.~M.,  2011, Astronomy \& Astrophysics, 527, A106

\bibitem[\protect\citeauthoryear{{Sob}, {Landman Bester}, {Smirnov}, {Kenyon}
  \& {Grobler}}{{Sob} et~al.}{2019}]{sob2019}
{Sob} U.,  {Landman Bester} H.,  {Smirnov} O.,  {Kenyon} J.,   {Grobler} T.,
  2019, arXiv e-prints, \href
  {https://ui.adsabs.harvard.edu/abs/2019arXiv191008136A} {p. arXiv:1910.08136}

\bibitem[\protect\citeauthoryear{{Spingola}, {McKean}, {Lee}, {Deller}  \&
  {Moldon}}{{Spingola} et~al.}{2019}]{spingola2019}
{Spingola} C.,  {McKean} J.~P.,  {Lee} M.,  {Deller} A.,   {Moldon} J.,  2019,
  \mn@doi [\mnras] {10.1093/mnras/sty3189}, \href
  {https://ui.adsabs.harvard.edu/abs/2019MNRAS.483.2125S} {483, 2125}

\bibitem[\protect\citeauthoryear{Taylor, Carilli  \& Perley}{Taylor
  et~al.}{1999}]{Syn1999}
Taylor G.~B.,  Carilli C.~L.,   Perley R.~A.,  eds, 1999, Synthesis Imaging in
  Radio Astronomy II, 1st edn.
 Astronomical Society of the Pacific Conference Series Vol. 180, Astronomical
  Society of the Pacific Conference Series

\bibitem[\protect\citeauthoryear{Thompson, Moran  \& George
  W.~Swenson}{Thompson et~al.}{2017}]{TMS2017}
Thompson A.~R.,  Moran J.~M.,   George W.~Swenson J.,  2017, Interferometry and
  Synthesis in Radio Astronomy, 3rd edn.
Springer Verlag

\bibitem[\protect\citeauthoryear{Trotta}{Trotta}{2008}]{trotta2008}
Trotta R.,  2008, Contemporary Physics, 49, 71

\bibitem[\protect\citeauthoryear{Whitney et~al.,}{Whitney
  et~al.}{2004}]{whitney2004}
Whitney A.~R.,  et~al., 2004, \mn@doi [Radio Science] {10.1029/2002RS002820},
  39

\bibitem[\protect\citeauthoryear{{Wielgus}, {Blackburn}, {Issaoun}, {Janssen},
  {Johnson}  \& {Koay}}{{Wielgus} et~al.}{2019}]{wielgus2019}
{Wielgus} M.,  {Blackburn} L.,  {Issaoun} S.,  {Janssen} M.,  {Johnson} M.,
  {Koay} J.,  2019, {EHT Memo Series, 2019-CE-02}

\bibitem[\protect\citeauthoryear{{van Bemmel}, {Small}, {Kettenis}, {Szomoru},
  {Moellenbrock}  \& {Janssen}}{{van Bemmel} et~al.}{2019}]{ilse2019}
{van Bemmel} I.,  {Small} D.,  {Kettenis} M.,  {Szomoru} A.,  {Moellenbrock}
  G.,   {Janssen} M.,  2019, arXiv e-prints, \href
  {https://ui.adsabs.harvard.edu/abs/2019arXiv190411747V} {p. arXiv:1904.11747}

\bibitem[\protect\citeauthoryear{{van Langevelde} et~al.,}{{van Langevelde}
  et~al.}{2019}]{huib2019}
{van Langevelde} H.,  et~al., 2019, arXiv e-prints, \href
  {https://ui.adsabs.harvard.edu/\#abs/2019arXiv190107804V} {p.
  arXiv:1901.07804}

\makeatother
\end{thebibliography}

\bsp	
\label{lastpage}
\end{document}